\documentclass[preprint,3p]{elsarticle}

\usepackage[colorlinks]{hyperref}
\usepackage[table]{xcolor}
\usepackage{placeins}
\usepackage{amsmath}
\usepackage{booktabs}
\usepackage{caption}
\usepackage{adjustbox}
\usepackage{amsfonts}
\usepackage{subfig}
\usepackage[capitalize]{cleveref}

\newcommand{\change}[1]{{\color{black}#1}}

\journal{arXiv}

\begin{document}

\begin{frontmatter}

\title{Using causal inference and Bayesian statistics to explain the capability of a test suite in exposing software faults}

\author[auth]{Alireza~Aghamohammadi}
\ead{aaghamohammadi@ce.sharif.edu}

\author[auth]{Seyed-Hassan Mirian-Hosseinabadi\corref{cor}}
\ead{hmirian@sharif.edu}

\cortext[cor]{Corresponding author}
\address[auth]{A. Aghamohammadi and S.-H. Mirian-Hosseinabadi are with the Department of Computer Engineering, Sharif University of Technology, Tehran, Iran.}

\begin{abstract}
Test effectiveness refers to the capability of a test suite in exposing faults in software.  
 It is crucial to be aware of factors that influence this capability.
 We aim at inferring the causal relationship between the two factors (i.e., \textit{Cover}/\textit{Exec}) and the capability of a test suite to expose and discover faults in software. \textit{Cover} refers to the number of distinct test cases covering the statement and \textit{Exec} equals the number of times a test suite executes a statement.
We analyzed {459166} software faults from {12} Java programs. Bayesian statistics along with the back-door criterion was exploited for the purpose of causal inference. 
Furthermore, we examined the common pitfall measuring association, the mixture of causal and noncausal relationships, instead of causal association. 
The results show that \textit{Cover} is of more causal association as against \textit{Exec}, and the causal association and noncausal one for those variables are statistically different.
Software developers could exploit the results to design and write more effective test cases, which  lead to discovering more bugs hidden in software.

\end{abstract}


\end{frontmatter}

\section{Introduction}\label{sec:Introcution}
There is no doubt that test efficacy, the degree to which a test suite exposes the latent faults embedded in software, is of importance~\citep{Aghamohammadi2020,Aghamohammadi2021,Grano2019,Holt2014,Inozemtseva2014,Mahdieh2020}\@. 
Apart from simple software programs or the usage of formal analysis, it is impossible or impractical to know the faults hidden in the software a priori. One way to address this issue and operationalize the concept of test efficacy is mutation testing~\citep{Grano2019,Li2009}\@. Mutation testing uses \textit{mutation operators} (syntactic rules to transform the source code) and generates artificial faults, mutants~\citep{Ammann2016,Chekam2017,Jia2011,Madeyski2010,Offutt2011,Papadakis2019,Pizzoleto2019}\@. 
An example of a mutant would be changing \texttt{x < 10} to \texttt{x > 10} in source code.

A mutant is run against the test suite and classified as \textit{alive} if all the test cases pass. If there is a test case that leads to a failure, the mutant is classified as \textit{killed}. 
Quality assurance experts try to write test cases that kill as many mutants as possible.
The number of killed mutants divided by the total number of \textit{non-equivalent mutants} is a quantity (i.e., \textit{mutation score}) that represents the test suite effectiveness. Equivalent mutants are the ones that have the same output as the original program~\citep{DelgadoPerez2020,Kintis2015,Offutt1996}. 
This study uses mutants as a substitute for faults. 

Both practitioners and researchers should be aware of factors that directly or indirectly increase the capability of a test suite in exposing and discovering faults in software~\citep{Grano2019,Jiarpakdee2020,Zhang2019}.
Examples of those factors would be code quality metrics (e.g., lines of code, Cyclomatic complexity~\citep{McCabe1989}\@, and object oriented design metrics~\citep{Cruz2017}) or  dynamic metrics (e.g., the number of times each fault is executed and the number of distinct test cases covering faults~\citep{Zhang2019}).

Recently, the software engineering community has brought attention to explainable AI where the predictions of statistical models are interpreted through investigating the importance of each feature locally or globally~\citep{Jiarpakdee2020,Jiarpakdee2020b,Jiarpakdee2021,Rajapaksha2021, xai4sebook,Tantithamthavorn2021}. In global interpretation, the association between independent variables and the dependent variable is examined for the whole data set. On the other hand, in local interpretation it is defined per instance.

\begin{figure*}[ht]
	\centering
	\includegraphics[width=0.90\linewidth]{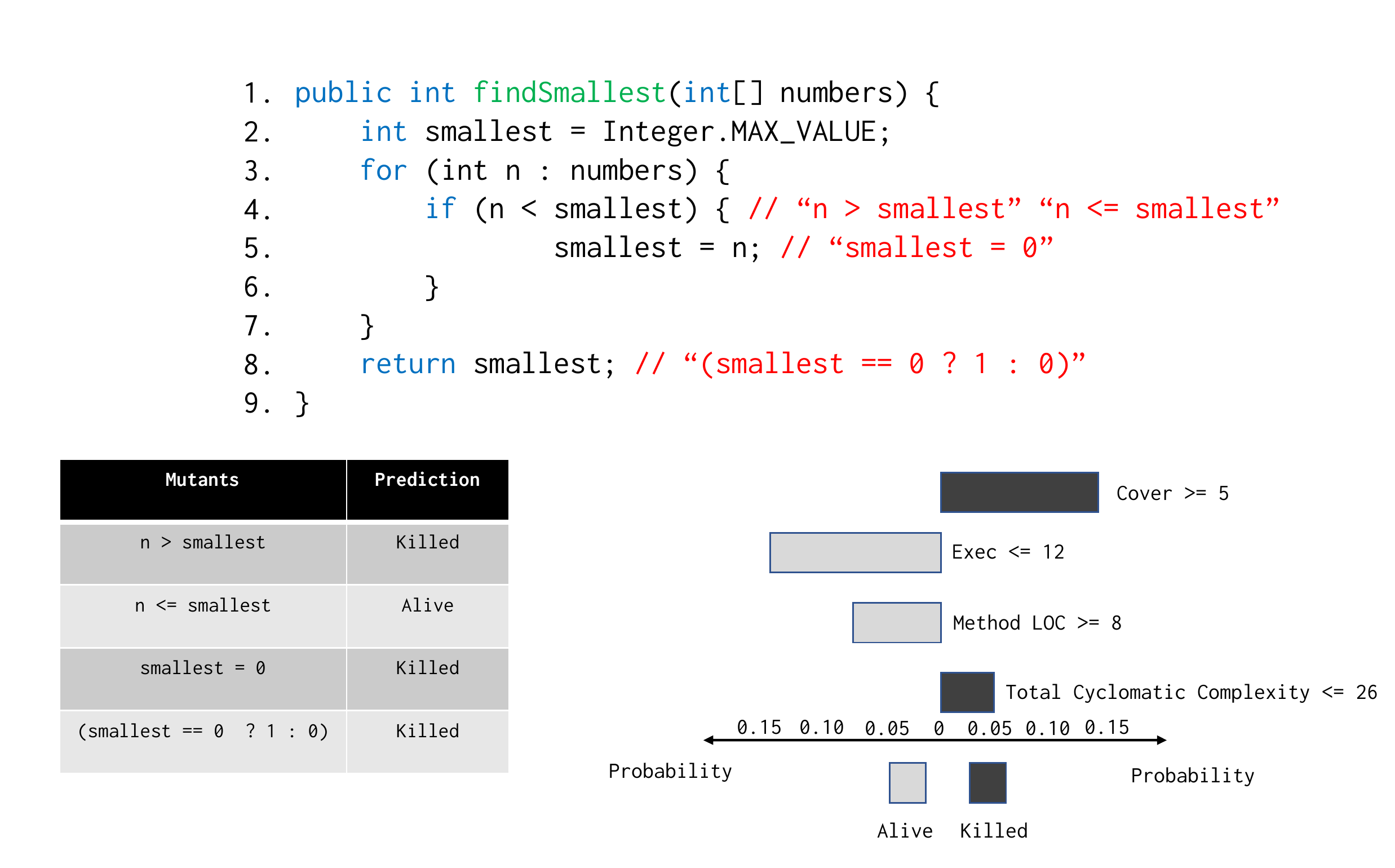}
	\caption{An example showing why drawing an inference from a statistical model is of importance}
	\label{fig:Motivating-example}
\end{figure*}

Take \cref{fig:Motivating-example} as an example where the part of a program accompanied by its mutants is demonstrated. 
This is an example showing why drawing an inference from a statistical model is of importance.
Suppose a statistical model is trained on the program. For that specific method (\texttt{findSmallest}), there are four mutants, three of which are predicted as killed. The mutant \texttt{n <= smallest} is  classified as alive and a developer wants to know why the statistical model classifies it as alive (explainable AI). If they were to interpret the independent variables, they would reach something similar to the bottom right corner of \cref{fig:Motivating-example} (e.g., using tools like LIME \footnote{\url{https://github.com/marcotcr/lime}}). For the sake of simplicity, assume there are four independent variables in this example:
the number of distinct tests covering the mutant, \textit{Cover}, the number of times each executed, \textit{Exec}, the method lines of code (LOC), and the total Cyclomatic complexity of the file in which the mutant is located.

The model believes that sufficient test cases cover the mutant (\texttt{Cover >= 5}), which means that there is no need to add new test cases for covering this part of the program.
On the other hand, since \textit{Exec} is less than or equal to 12, it suggests that a developer should increase \textit{Exec} to kill the mutant. Since the total Cyclomatic complexity of the file is low enough, it might suggest that there is no need to refactor the source code in order to reduce the Cyclomatic complexity value. However, the method LOC is greater than eight and the model thinks the method lines of code is large. A developer, for example, could use a built-in method provided by Java (e.g., \texttt{Collections.min}) to find the minimum value instead of implementing it from scratch, which reduces the method lines of code from eight to one.

However, the aforementioned interpretation is completely based on association, which is different from causation.
As for software engineers, they require to know to what extent changing the number of executions (\textit{Exec}) or the number of covering tests (\textit{Cover}) lead to empowering the capability of a test suite in exposing software faults. 
In this specific example, the mutant \texttt{n <= smallest} is classified as alive. In order to kill that mutant, the question is whether developers should write new tests to increase the number of distinct tests covering it (and at the same time increase the number of exeuctions), they should modify the existing tests to increase the number of times that \texttt{n <= smallest} is executed, or they should follow an strategy which combines the both approaches.

Causal inference fills this gap by providing the reason and solution altogether, guiding practitioners through not only why a variable should change but also how this change affects the outcome concerned, and quantifying the causal effect of each change. One way to provide causal claims is using Randomized Controlled Trial (RCT). 
Unfortunately, RCT for drawing causal inferences is not always feasible~\citep{ Baah2010,Bai2016, Kucuk2021,McElreath2020,Neal2021,Shu2013}\@. Randomization does not guarantee whether all the contributing factors are balanced in the treatment and control group. It may be impractical, infeasible, or even not possible in nature to control confounding factors. Another problem with RCT lies in generalizability.  RCT tends to be conducted in a lab experiment. The results, therefore, might not hold in a field study or real-world settings. \textit{Observational studies}, on the other hand, measure the variables of interest and make an inference based on the sample drawn from the population concerned~\citep{Nichols2007,Shull2008}.

This paper poses the question of which factor (i.e., \textit{Cover} or \textit{Exec}) and to what extent \textit{causally}, not simple correlation, influences the capability of a test suite in exposing software faults in an observational study. 
These two metrics are the two most important metrics in assessing test effectiveness~\citep{Mao2019,Zhang2019,Zhang2020}\@.

To this end, we analyzed {459166} mutants from {12} open source Java projects. In order to make a causal inference, we created causal Directed Acyclic Graph (DAG) and exploited the back-door criterion~\citep{Gelman2011,Guo2020,McElreath2020,Neal2021,Pearl2009}. The causal DAG is a graph where each node represents the variables (dependent or independent) and each edge indicates the causal relationship between two variables.
The back-door criterion offers variables on which we could control for, the \textit{sufficient adjustment set}~\citep{Neal2021}. Controlling for these variables blocks the \textit{back-door paths}, which are noncausal paths from the variable of interest, $T$, to the dependent variable, $Y$, in the specified causal DAG and have incoming edges to $T$.

We exploited Bayesian statistics as against Frequentist statistics~\citep{vandeSchoot2021}\@. Frequentist statistics is not a suitable option for causal inference since this study is not a RCT experiment. 
In the Bayesian view, it is feasible to condition on unknown parameters (i.e., the parameters we would like to make inferences). This feature is pivotal to  draw causal inferences as paving the way to control for the sufficient adjustment set.
Furthermore, 
Bayesian statistics gives us a distinct advantage over Frequentist statistics as providing probability distributions of interest, as against point estimates~\citep{Furia2019,Torkar2021}\@. It maps a prior belief (i.e., a prior distribution) to the posterior distribution of interest using observed data. 
This mapping is accomplished through Bayes' theorem~\citep{Cornfield1967}\@. In this study, the probability distribution function of individual mutant results (i.e., killed or alive) is the distribution of interest.
We followed a fully Bayesian approach, avoiding Frequentist terms like p-value or effect size as suggested by many Bayesian statisticians~\citep{Correll2020,Halsey2019,McElreath2020,Wasserstein2016}.
In this article, we make the following contributions:

\begin{itemize}
	\item We show, even in the simplest form where there are few independent variables, there is a pitfall which many researchers fall into, measuring noncausal association instead of the causal one. In the complex models with tens of or even hundreds of independent variables, it is more likely to make inaccurate inference.
	\item We measure the causal relationships from \textit{Cover} and \textit{Exec} to the mutant execution results (i.e., killed or alive) in an observational study, which comprises {459166} mutants from {12} open source Java projects. The causal relationships between independent variables and outcome of interest could provide researchers with an accurate impact of those on the outcome.
	\item We demonstrate that the causal and noncausal association of \textit{Exec} differ statistically, meaning researchers should be highly cautious when interpreting a statistical model.
	\item We provide all the source code and data for other scientists, enabling them to replicate this study\footnote{\url{https://github.com/aaghamohammadi/Causal-Test-Efficacy}}.
\end{itemize}

The rest of the article is structured as follows.
\cref{sec:Related} examines the related work.
\cref{sec:Background} explains the preliminaries and necessary background information for understanding the paper, namely Bayesian statistics and the back-door criterion. In \cref{sec:Methodology}, we elaborate upon our methodology used to answer the research questions regarding the influence of \textit{Cover} and \textit{Exec} on mutants. \cref{sec:Results} presents the findings and answers the research questions. In \cref{sec:Discussion}, we discuss the limitations of the suggested approach, the implication of this research, and the impact on developers in general. Afterward, the threats to the validity of the paper is outlined in \cref{sec:Threats} and finally we conclude the paper in \cref{sec:Conclusions}.

\section{Related Work}\label{sec:Related}
Little research in software engineering has employed Bayesian statistics~\citep{Ernst2018, Furia2019,Furia2021, Scholz2020,Torkar2020, Torkar2021}\@. \citet{Ernst2018} applied multi-level GLM to the problem of finding the average Coupling Between Objects (CBO) metrics for a heterogeneous data set. Multi-level GLM is suitable for this issue since it uses partial pooling, taking advantages of both local and global information. Using the multi-level GLM declined the prediction error by {50}\%.

\citet{Furia2019} thoroughly studied, examined, and demonstrated the merits of the Bayesian data analysis. They reanalyzed two studies which have originally exploited Frequentist statistics~\citep{Ceccato2015}. The first study is related to automatically generated test cases and the second one compares eight programming languages in terms of performance~\citep{Nanz2015}\@. They concluded that Frequentist viewpoint is unintuitive and hard-to-interpret, and it should be abandoned and replaced by a more rigor, sound, and intuitive approach (i.e., Bayesian perspective).

\citet{Torkar2021} claimed that Bayesian data analysis could be combined with  cumulative prospect theory, providing practical significance. 
They reanalyzed a case study, comparing exploratory testing and document-based testing~\citep{Afzal2014}\@.
Practical significance is of help in making decisions at a managerial level. Bayesian statistics was served since it helps to infer, estimate, and interpret the risk of each decision in terms of probability distribution. 

\citet{Scholz2020} used Bayesian statistics and the causal DAG to solve the problem of the defect prediction task using past faults. Similar to our work, they used multi-level GLMs (i.e., indexing the selected projects) and the back-door criterion. As opposed to our work, they focused on the evaluation of their suggested algorithm for fault prediction task and their aim was not comparison between causal and noncausal association. The problem context is also another difference, focusing on test effectiveness compared to defect prediction.

Using mutants as a proxy for faults in order to assess test effectiveness has a long history~\citep{Aghamohammadi2020,Aghamohammadi2021,Chekam2019,Cruz2017,Dallilo2019,DuqueTorres2020,Papadakis2018,Zhang2019}. In our previous work, we proposed a new coverage metric named Statement Frequency Coverage to assess the test suite efficacy~\citep{Aghamohammadi2021}. We analyzed {22} Python programs and existing test cases and exploited mutants as a substitute for faults. We measured the proposed code coverage, statement coverage, branch coverage, and mutation score for specified projects, showing the correlation between the proposed code coverage and mutation score is stronger than that of statement and branch coverage.

\section{Background}\label{sec:Background}
We briefly explain Bayesian statistics and causal inference to acquire information for understanding this article.
\subsection{Bayesian statistics}
In a Bayesian world, two types of uncertainties exist~\citep{Gelman2013, OHagan2004}\@: \textit{aleatory} and \textit{epistemic}. These two uncertainties differ in a key way. The aleatory uncertainty arises from random effects, e.g., random measurement errors. The epistemic uncertainty is due to the lack of knowledge, reduced by gathering data. The more data with respect to the unknown parameter of interest we have, the less epistemic uncertainty will be. 

Bayesian statistics models the uncertainties using Bayes' theorem~\citep{Cornfield1967, Gelman2013}\@:
\begin{equation}\label{eq:Bayes-rule}
\textrm{P}\left(\theta \mid y\right) = \frac{\textrm{P}\left(y \mid \theta\right) \textrm{P}\left(\theta\right)}{\textrm{P}\left(y\right)}
\end{equation}
In \cref{eq:Bayes-rule}, $\theta$ and $y $ refer to the unknown parameter of interest and data, respectively.  Bayes theorem provides information about the unkown parameter using known data and maps our prior belief about $\theta$, $\textrm{P}\left(\theta\right)$, to the posterior distribution, $\textrm{P}\left(\theta \mid y\right)$, having seen the data. This transformation is accomplished using the likelihood function, $\textrm{P}\left(y \mid \theta\right)$. The term $\textrm{P}\left(y\right)$ makes the posterior distribution a well-defined probability density function.

As an illustrative example, let $\theta$ be the unknown mutation score for a given project. Suppose we do not have any background information regarding $\theta$ and assign a uniform distribution (\textit{uninformative} prior) to the prior, $\textrm{P}\left(\theta\right)$~\citep{Gelman2013}\@. We draw {100} mutants from the total population, of which {70} mutants are classified as killed. Assume a mutant execution result is independent from one another (i.e., Binomial likelihood); therefore, the updated version of $\theta$, $\mathrm{P}\left(\theta \mid y\right)$, becomes the $\textrm{Beta}$ distribution with the parameters ${71}$ and ${31}$. Once the posterior distribution is calculated, we can determine the quantities of interest such as the expected value (i.e., $\mathrm{E}\left(\theta \mid y\right) = \frac{70}{70 + 30} = {0.7}$), which is the point estimate of mutation score. The point estimate is similar to the Frequentist view point. However, in the Bayesian world we can infer probabilistic characteristics like {95}\% \textit{credible interval} (i.e., $\textrm{P}\left({0.603} < \theta < {0.781} \mid y\right) = {0.95}$), meaning with the probability of {95}\% the mutation score lies between $0.603$ and $0.781$.

\subsection{Causal inference}
A variable $X$ has a causal effect on a variable $Y$ provided that $Y$ is able to change in reaction to modifications in $X$~\citep{Guo2020, McElreath2020, Neal2021, Pearl2009}\@. The causal relationships between variables are generally modeled by causal DAG where nodes are the variables and edges indicate the causal association. For example if $X$ has a causal effect on $Y$, there is a direct edge from $X$ to $Y$ (i.e., $X \rightarrow Y$).

Causal DAG comprises three kinds of building blocks~\citep{Guo2020,McElreath2020,Neal2021}\@:
\begin{itemize}
\item \textit{Chain/Pipe}: For three variables $X_1$, $X_2$, and $X_3$ we have a chain/pipe provided that $X_1 \rightarrow X_2 \rightarrow X_3$. As a result, there is an association, statistical dependence, between $X_1$ and $X_3$ in the chain/pipe. Conditioning on $X_2$ makes $X_1$ and $X_3$ independent.
\item \textit{Fork}: For three variables $X_1$, $X_2$, and $X_3$ there exists a fork if $X_1 \leftarrow X_2 \rightarrow X_3$. 
As a result, $X_2$ is a common cause of $X_1$ and $X_3$.
In a fork, there is a back-door path (i.e., a noncausal association) from $X_1$ to $X_3$ through $X_2$. Conditioning on $X_2$ blocks the path and makes $X_1$ and $X_3$ independent.

\item \textit{Immorality/Collider}: Three variables $X_1$, $X_2$, and $X_3$ form a collider if $X_1 \rightarrow X_2 \leftarrow X_3$. By default, there is no open path from $X_1$ to $X_3$. Conditioning on $X_2$ makes the path open and creates a noncausal association between $X_1$ and $X_3$.
\end{itemize}

Exploiting these three types of building blocks, the back-door criterion provides us with the variables, the sufficient adjustment set, we should control for in order to block noncausal paths from the variable of interest to the outcome~\citep{Guo2020, McElreath2020, Neal2021}\@.

\begin{figure}[ht]
    \centering
    \includegraphics[width=0.7\linewidth]{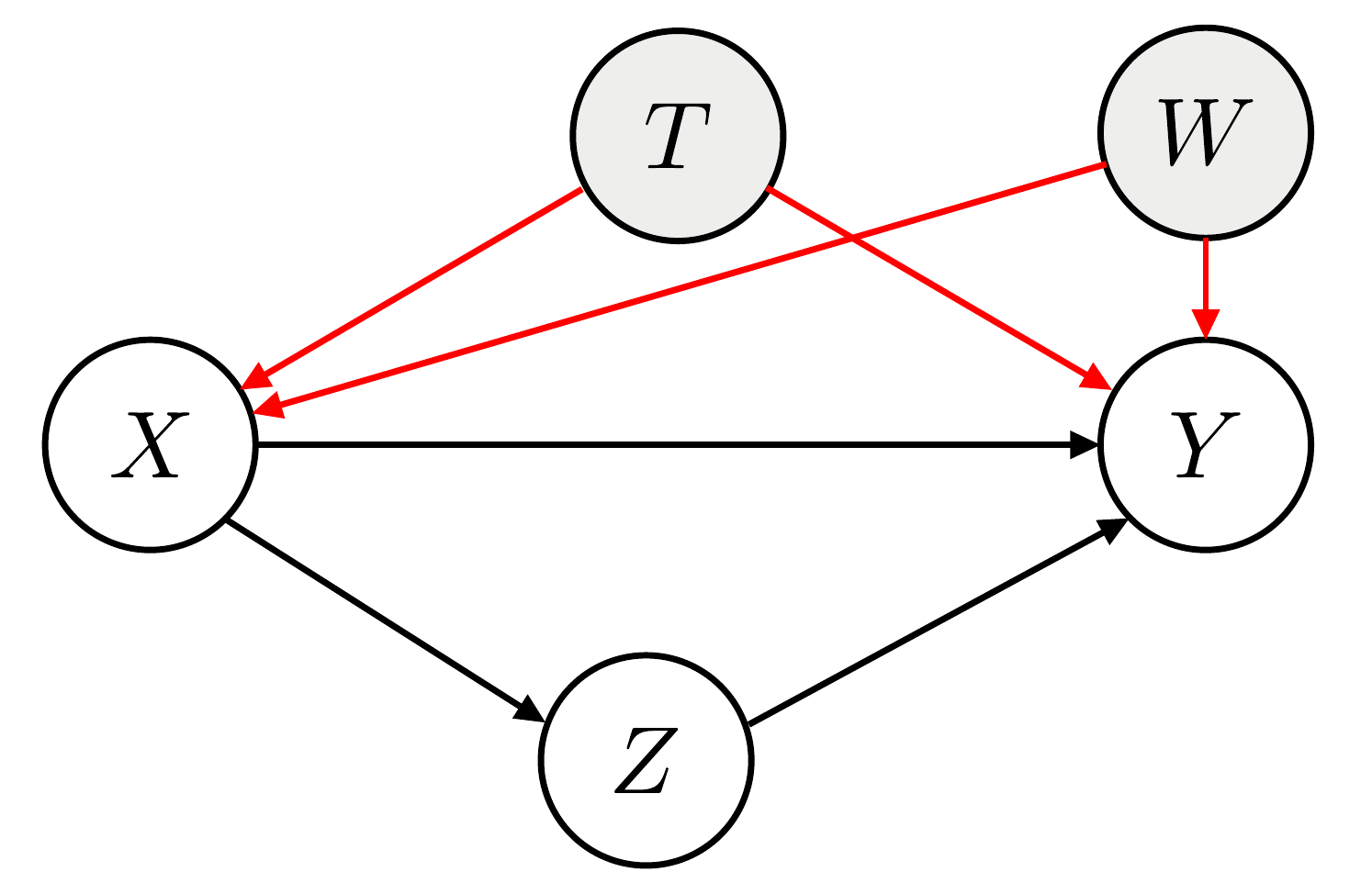}
    \caption{An example for the back-door criterion}
    \label{fig:backdoor}
\end{figure}

To better understand the back-door criterion, consider the example illustrated in \cref{fig:backdoor} where we would like to determine the causal effect of $X$ on $Y$. 
In this figure $X$, $T$, $W$, and $Z$ are independent variables and $Y$ is a dependent variable.
There are four paths from $X$ to $Y$:
\begin{enumerate}
\item $X \rightarrow Y$
\item $X \leftarrow T \rightarrow Y$
\item $X \leftarrow W \rightarrow Y$
\item $X \rightarrow Z \rightarrow Y$
\end{enumerate}

Of those, the paths $X \leftarrow T \rightarrow Y$ and $X \leftarrow W \rightarrow Y$ are noncausal and they should be blocked by conditioning on $T$ and $W$. The paths $X \rightarrow Y$ and $X \rightarrow Z \rightarrow Y$ are causal and we should not control for the variable $Z$ if the causal association from $X$ to $Y$ is of desire.

In this scenario, Bayesian statistics is of help by providing the complete probability distributions for a variable concerened, supporting conditioning on the sufficient adjustment set, and generating intuitive-to-interpret results~\citep{McElreath2020,Neal2021}\@.

\section{Methodology}\label{sec:Methodology}
In this paper, we draw causal inferences from an observational study. First, we describe selected subjects, their characteristics, and the independent/dependent variables. Then, the research questions along with their objectives are posed. The first research question shows the common mistake many fall into, measuring noncausal relationships. The second research question demonstrates how causal association should be determined using the back-door criterion. The third one aims at answering counterfactual conditionals. The final research question indicates another common pitfall many committed. After that, the procedure of exploratory analysis for answering the research questions is articulated. Finally, we explain evaluation metrics used in the study.
\subsection{Subjects}
We initially started on four Java projects from a study by \citet{Zhang2020}\@, namely \textsc{la4j}, \textsc{lang}, \textsc{msg}, and \textsc{wire}. These four projects are also used in other published studies \citep{Mao2019,Zhang2016,Zhang2019}\@. To increase the diversity of selected projects, additional eight subjects were randomly selected from research by \citet{Mao2019}. Each project contains developer-written tests and mutant execution results (i.e., alive or killed as a dependent variable). Mutants were generated using PIT, a well-known Java mutation testing tool \citep{Coles2016}\@. 

Two independent variables were measured for a mutant:
\begin{itemize}
    \item \textit{Exec}: an integer (e.g., 0, 1, 2, 3, 4, $\dots$) indicating the number of times the test suite runs the mutant.
    \item \textit{Cover}: an integer (e.g., 0, 1, 2, 3, 4, $\dots$) representing the number of distinct test cases that execute the mutant.
\end{itemize}
 
\cref{tab:data} shows the statistics of the collected subjects. 
Subjects vary in size from {4526} mutants, \textsc{argparse4j}, to {127929} mutants, \textsc{opennlp}. Above 30 developers, for example, contribute to \textsc{opennlp} and it has more than {35} releases. The projects come from different domains such as natural language processing (\textsc{opennlp}), assertions for testing (\textsc{assert4j-core}), date time (\textsc{joda-time}), and linear algebra (\textsc{la4j}), to mention but a few. 

Unsurprisingly the distributions of \textit{Exec} and \textit{Cover} variables are right-skewed (see skewness in \cref{tab:data}). On the one hand, the medians of \textit{Exec} for most projects are below {25} except for \textsc{la4j} and \textsc{msg}. On the other hand, there are few extreme cases where mutants are executed repeatedly (e.g., {2.5} billion or more for \textsc{opennlp}). The same is true for the \textit{Cover} variable, except for \textsc{recast4j} which is not skewed. All of the selected projects have a median fewer than {10}. However, there are some cases, e.g. \textsc{assert4j-core} where a mutant is executed by {9835} different test cases.

\begin{table*}[htbp]
\centering
\caption{Statistics of the collected projects}
\label{tab:data}
\begin{adjustbox}{width=\linewidth,center}
\begin{tabular}{@{}llllllllll@{}}
\toprule
\textbf{Subject}&   \textbf{\# Mutants}&    \textbf{MS}&  \textbf{Variable}&   \textbf{Min}&   \textbf{Q1}&   \textbf{Median}&   \textbf{Q3}&    \textbf{Max}&   \textbf{Skewness}\\
\midrule
\textsc{argparse4j}&   {4526}&   {0.72}&Exec&   {0}&   {2}&   {14}&  {64}& {1850}&   {4.47}\\
\addlinespace
&   &   &Cover&   {0}&   {1}&   {4}&   {19}& {121}&   {2.52}\\
\midrule
\textsc{assertj-core}&   {26520}&   {0.83}&Exec&   {0}&   {4}&   {12}&   {66}&   {8778402}&   {30.90}\\
\addlinespace
&   &   &Cover&   {0}&   {2}&   {5}&   {18}&   {9835}&   {8.81}\\
\midrule
\textsc{fess}&   {154187}&   {0.04}&   Exec&   {0}&   {0}&   {0}&   {0}&   {19857948}&   {36.73}\\
\addlinespace
&   &   &Cover&   {0}&   {0}&   {0}&   {0}&   {122}&   {55.63}\\
\midrule
\textsc{joda-time}&   {35748}&   {0.74}&Exec&   {0}&   {3}&   {22}&   {306}&   {218487062}&   {30.88}\\
\addlinespace
&   &   &Cover&   {0}&   {1}&   {5}&   {33}&   {3453}&   {6.41}\\
\midrule
\textsc{la4j}&   {13708}&   {0.64}&Exec&   {0}&   {0}&   {60}&   {779}&   {876609116}&   {17.70}\\
\addlinespace
&   &   &Cover&   {0}&   {0}&   {5}&   {26}&   {686}&   {4.31}\\
\midrule
\textsc{lang}&   {20238}&   {0.74}&Exec&   {0}&   {7}&   {19}&   {89}&   {17900385}&   {25.07}\\
\addlinespace
&   &   &Cover&   {0}&   {2}&   {3}&   {7}&   {264}&   {5.14}\\
\midrule
\textsc{msg}&   {7101}&   {0.52}&Exec&   {0}&   {0}&   {74}&   {5589}&   {13222733}&   {13.71}\\
\addlinespace
&   &   &Cover&   {0}&   {0}&   {6}&   {107}&   {1117}&   {2.40}\\
\midrule
\textsc{nodebox}&   {38334}&   {0.26}&Exec&   {0}&   {0}&   {0}&   {4}&   {327864}&   {55.02}\\
\addlinespace
&   &   &Cover&   {0}&   {0}&   {0}&   {1}&   {149}&   {6.31}\\
\midrule
\textsc{opennlp}&   {127929}&   {0.44}&Exec&   {0}&   {0}&   {10}&   {970}&   {2569352493}&    {32.30}\\
\addlinespace
&   &   &Cover&   {0}&   {0}&   {1}&   {7}&   {231}&   {4.82}\\
\midrule
\textsc{recast4j}&   {13685}&   {0.59}&Exec&   {0}&   {82}&   {30432}&   {718121}&   {1218552274}&   {13.37}\\
\addlinespace
&   &   &Cover&   {0}&   {2}&   {7}&   {9}&   {12}&   {-0.17}\\
\midrule
\textsc{uaa}&   {11391}&   {0.51}&Exec&   {0}&   {0}&   {4}&   {22}&   {7734}&   {15.03}\\
\addlinespace
&   &   &Cover&   {0}&   {0}&   {2}&   {9}&   {11}&   {3.66}\\
\midrule
\textsc{wire}&   {5799}&   {0.29}&Exec&   {0}&   {0}&   {0}&   {6}&   {13701}&   {8.77}\\
\addlinespace
&   &   &Cover&   {0}&   {0}&   {0}&   {2}&   {68}&   {2.72}\\
\bottomrule
\multicolumn{10}{@{}p{\linewidth}@{}}{MS refers to mutation score. Q1 and Q3 are the first and the third quartile respectively.}\\
\end{tabular}
\end{adjustbox}
\end{table*}

\textbf{Data Preprocessing.}
The independent variables \textit{Exec} and \textit{Cover} are transformed in order to be easier for us to opt the priors and interpret the statistical models for transformed data. We used two transformations for each project. First, we applied the $x \rightarrow \log(x + 1)$ transformation to reduce the right skewness. After that, we standardized these two variables for a project, making them centered around zero. As such, the averages of transformed data, with respect to each project, for \textit{Exec} and \textit{Cover} become zero.

\subsection{Research questions}

\begin{description}
\item[RQ1:] What is the noncausal relationship between \textit{Exec} and mutant execution results?

RQ1 is meant to measure the  association, statistical dependence, between the explanatory variable, \textit{Exec}, and the outcome of interest, ignoring the effect of confounding variables. The aim is to demonstrate the common pitfall many fall into, calculating statistical association which does not arise from causal relationship. We argue that ignoring other variables thoughtlessly overestimates the statistical association between two variables.

\item[RQ2:] What is the causal relationship between \textit{Exec} and mutant execution results?

RQ2 is intended to estimate the causal association between \textit{Exec} and the outcome concerned, considering the impact of \textit{Cover}. The back-door criterion is exploited for this purpose. The main hypothesis of RQ2 is that the causal association between \textit{Exec} and the outcome is significantly less than the noncausal relationship measured in RQ1.

\item[RQ3:] Having known \textit{Cover}, how much additional information \textit{Exec} provides in knowing mutant execution results?

RQ3 explores the implication of manipulating \textit{Exec} on the mutant, controlling the variable \textit{Cover}. To this end, \textit{Counterfactual plots} are served (see \citet{McElreath2020}) to answer \textit{what-if} questions such as what would happen in case of increasing \textit{Exec} to an imaginary value, holding \textit{Cover} constant.

\item[RQ4:] What is the  causal relationship between \textit{Cover} and the outcome?

In RQ4 we argue that to measure the  causal relationship between \textit{Cover} and the mutant execution results, one should ignore the impact of \textit{Exec} on the outcome. As such, controlling \textit{Exec} could distort the association between \textit{Cover} and mutants.
\end{description}

\subsection{Exploratory analysis}
\paragraph{RQ1: Noncausal relationship between \textit{Exec} and the outcome}\mbox{}\\
To answer RQ1, we created a statistical model where mutant execution results (i.e., alive or killed) is regarded as the dependent variable and \textit{Exec} as the explanatory variable. The following are the model details.

\begin{subequations}
\begin{align}
    M_i &\sim \textrm{Bernoulli}\left(\theta_i\right)\label{eq:RQ1-m-1}\\
    \textrm{logit}\left(\theta_i\right) &= \alpha_{\mathrm{PRJ}\left[i\right]} + \beta_{\mathrm{PRJ}\left[i\right]} \times \textrm{Exec}_{i}\label{eq:RQ1-m-2}\\
    \alpha_{\mathrm{PRJ}\left[i\right]} &\sim \textrm{Normal}\left(0,1\right)\label{eq:RQ1-m-3}\\
    \beta_{\mathrm{PRJ}\left[i\right]} &\sim \textrm{Log{-}Normal}\left(0,1\right)\label{eq:RQ1-m-4}
\end{align}
\end{subequations}

A dichotomous variable $M_i$ refers to the mutant outcome (i.e., {1} for killed and {0} for alive) and $\textrm{Exec}_{i}$ is the corresponding execution number. \cref{eq:RQ1-m-1} is the Bernoulli likelihood function (i.e., $\mathrm{P}\left(M_i \mid \theta_i\right)$) with the parameter $\theta_i$, the probability of a mutant being killed. As $\theta_i \in \left[0,1\right]$, we exploited the logit, log-odds function ($\mathrm{logit}\left(\theta_i\right) = \log \frac{\theta_i}{1 - \theta_i}$), on the left-hand side of \cref{eq:RQ1-m-2}. It is a common choice for the unknown parameter of Bernoulli (or Binomial in general) since it maps $\left[0, 1\right]$ to $\mathbb{R}$~\citep{Gelman2013,McElreath2020}\@. 

On the right-hand side of \cref{eq:RQ1-m-2} lies the core of the multi-level generalized linear model.  It connects the unknown parameter $\theta_i$ to the known explanatory variable $\textrm{Exec}_{i}$, where $\alpha_{\mathrm{PRJ}\left[i\right]}$ and $\beta_{\mathrm{PRJ}\left[i\right]}$ become the new unknown parameters. It is multi-level since intercepts and slopes are varying per project. Each project has unique range of \textit{Exec}, thereby having different influences on mutants. Whereas, the creation of multiple distinct models neglects the latent knowledge we can harness and transfer from one project to another. Multi-level GLM solves this dilemma by varying intercepts and slopes. This technique is known in literature as partial-pooling\citep{Ernst2018, Gelman2013, McElreath2020}\@.

\cref{eq:RQ1-m-3} and \cref{eq:RQ1-m-4} are the priors for the parameters. 
We follow the \textit{weakly-informative priors} for all our statistical models~\citep{Gelman2013}\@. The weakly-informative priors are the ones ensuring that the range of variables is plausible before observing the data~\citep{McElreath2020}\@. 

Because the data is centered around zero, the expected value of $\alpha_{\mathrm{PRJ}\left[i\right]}$ equals $\mathrm{E}\left(\textrm{logit}\left(\theta_i\right)\right)$, which is the mutation score on the logit scale. The mutation score of {50}\%, {90}\% and {10}\% has {0}, {2.2}, and {-2.2} logit values respectively. The prior defined in \cref{eq:RQ1-m-3} implies that before seeing the data we expect most of the projects (approximately 95\%) to have mutation score between {10}\% and {90}\%. Furthermore, for a given mutant, the more it is executed, the more plausible it is to be classified as killed. As a result, it is expected that $\beta_{\mathrm{PRJ}\left[i\right]} \ge 0$. \cref{eq:RQ1-m-4} represents this belief~\citep{McElreath2020}\@.

\begin{figure*}[ht]
    \centering
    \includegraphics[width=0.7\linewidth]{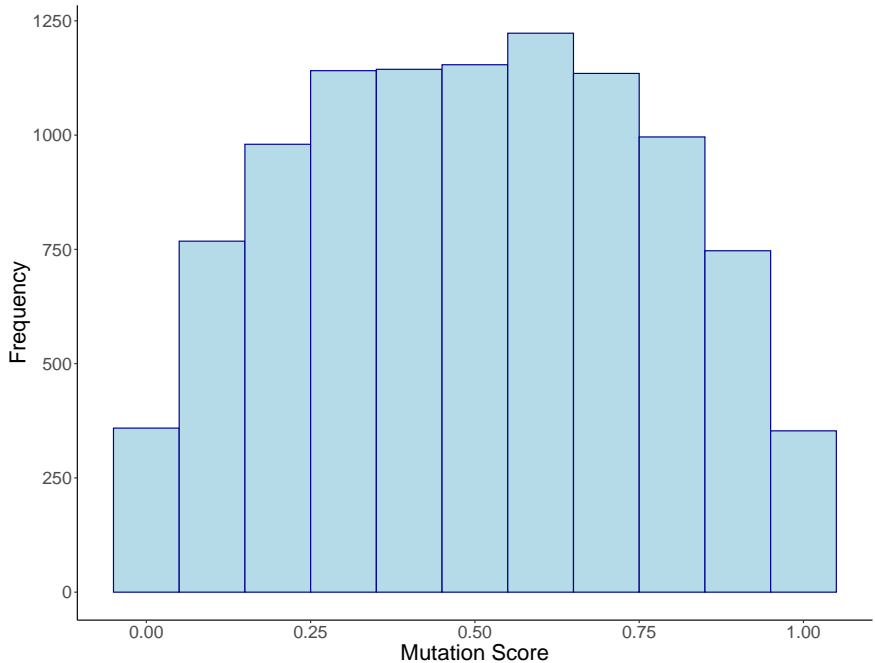}
    \caption{Prior predictive simulation}
    \label{fig:prior-predictive-checks}
\end{figure*}

To ensure that the chosen priors are reasonable, we used \textit{prior predictive simulations}. Prior predictive simulations are the way to predict the outcome of interest using just the priors (i.e., without data)~\citep{Furia2021,Gelman2013, McElreath2020,Scholz2020}\@. Mathematically, in this case, it is the same as $\mathrm{P}\left(M\right) = \int_{\Theta}^{}(\mathrm{P}\left(M,\theta\right )d\theta$ where $M$ is a mutant execution result. \cref{fig:prior-predictive-checks} demonstrates the frequency of mutation score (i.e., $\mathrm{E}\left(M\right)$) these priors suggest. As can be seen, the extreme mutation score, above {90}\% or below {10}\%, is less plausible (before seeing the data).

\begin{figure*}[ht]
    \centering
    \includegraphics[width=0.7\linewidth]{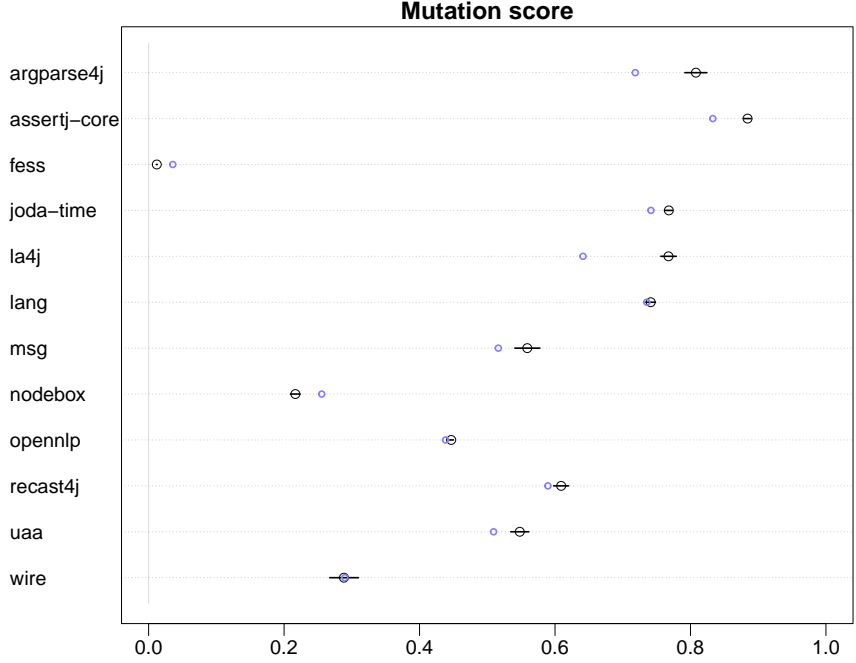}
    \caption{Posterior checks}
    \label{fig:posterior-checks}
\end{figure*}

In addition to prior predictive simulations, we applied posterior predictive checks~\citep{McElreath2020} for the mutation score of the projects. We would like to see whether the observed mutation score is a cogent realization from the posterior of average over $\theta$. \cref{fig:posterior-checks} shows the posterior predictive checks for the expected value of $\theta$ against actual mutation score. The dark circles are the predictions and the light ones are actual values. The intervals are the {95}\% credible intervals on the probability scale.
The output is reasonable as the mean Bayesian R-squared (the Bayesian version of R-squared~\citep{Gelman2019}) is ${0.98}$.

\paragraph{RQ2: Causal relationship between \textit{Exec} and the mutant execution results}\mbox{}\\

\begin{figure}[ht]
    \centering
    \includegraphics[width=0.7\linewidth]{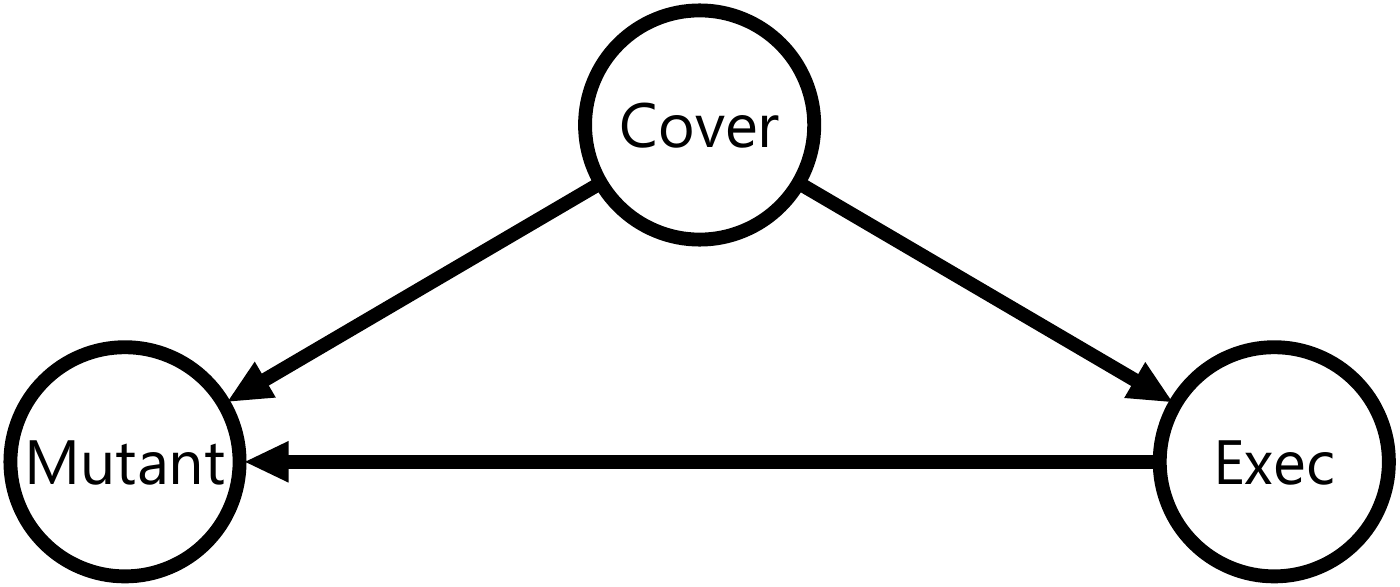}
    \caption{Causal DAG between \textit{Cover}, \textit{Exec}, and mutants}
    \label{fig:causal-dag}
\end{figure}

We use the causal DAG depicted in \cref{fig:causal-dag}. The arrow from \textit{Cover} to \textit{Exec} is due to the fact that increasing \textit{Cover} necessarily leads to an increase in \textit{Exec}. 
Note that the reverse is not correct. Increasing \textit{Exec} does not necessarily lead to an increase in \textit{Cover}.

The increase in \textit{Cover} leads to an increase in the odds of a mutant being killed. As a result, there is an arrow from \textit{Cover} to the mutant. With the same argument, there should be an arrow from \textit{Exec} to the mutant. 

One may hypothetically ask that why independent variables such as mutation operators, method LOC, Cyclomatic complexity, and so forth are not added to the causal DAG since there are correlations between them and the outcome. Variables like mutation operators do not have a causal effect on \textit{Exec} and \textit{Cover}; therefore, they do not create a back-door path. In other words, they are not confounding factors and should be omitted from the graph. If they add to the statistical model without in-depth investigation of the impact they impose, it could have side effects and distort the results.

Since there is a back-door path in \cref{fig:causal-dag} ($\textrm{Mutant} \leftarrow \textrm{Cover} \rightarrow \textrm{Exec}$), \textit{Cover} becomes a fork. 
Therefore, it requires us to add that explanatory variable to the statistical model in order to measure the  causal association between \textit{Exec} and the outcome concerned. The subsequent multi-level GLM is intended to accomplish this.

\begin{subequations}
\begin{align}
    M_i &\sim \textrm{Bernoulli}\left(\theta_i\right)\label{eq:RQ2-m-1}\\
    \textrm{logit}\left(\theta_i\right) &= \alpha_{\mathrm{PRJ}\left[i\right]} + \beta_{\mathrm{PRJ}\left[i\right]} \times \textrm{Exec}_{i} + \gamma_{\mathrm{PRJ}\left[i\right]} \times \textrm{Cover}_{i}\label{eq:RQ2-m-2}\\
    \alpha_{\mathrm{PRJ}\left[i\right]} &\sim \textrm{Normal}\left(0,1\right)\label{eq:RQ2-m-3}\\
    \beta_{\mathrm{PRJ}\left[i\right]} &\sim \textrm{Log{-}Normal}\left(0,1\right)\label{eq:RQ2-m-4}\\
    \gamma_{\mathrm{PRJ}\left[i\right]} &\sim \textrm{Log{-}Normal}\left(0,1\right)\label{eq:RQ2-m-5}
\end{align}
\end{subequations}

Two differences arise out of the \textit{Cover} presence, as against the specified model in RQ1. First, \cref{eq:RQ2-m-2} includes $\gamma_{\mathrm{PRJ}\left[i\right]} \times \textrm{Cover}_{i}$ to the right-hand side of the logit function. It causes $M_i$ to be conditioned on \textit{Cover} when computing the probability of $\mathrm{P}\left(M_i \mid \theta_i\right)$. As a result, it blocks the back-door path from \textit{Exec} to the mutant. Second, \cref{eq:RQ2-m-5} defines the prior for the $\gamma_{\mathrm{PRJ}\left[i\right]}$ coefficient. The same argument, used in RQ1 for $\beta_{\mathrm{PRJ}\left[i\right]}$, applies here for the usage of Log-Normal. To save space, we avoid reporting prior predictive simulations and posterior checks for this statistical model, yet it is available on the online appendix (see contributions).
Note that \cref{eq:RQ2-m-1} and \cref{eq:RQ1-m-1} are the same since the outcome $M_i$ is dichotomous (i.e., 1 for killed and 0 for alive) and follows the Bernoulli likelihood function with the parameter $\theta_i$, which defines the probability of a mutant being killed.

\paragraph{RQ3: Counterfactual conditionals}\mbox{}\\
To answer counterfactual propositions, we need a model to simultaneously estimate the effect of \textit{Cover} on \textit{Exec} as well as calculating the causal association between \textit{Exec} and the outcome~\citep{McElreath2020,Rips2010}\@. As such, the statistical model should include two GLMs. The following articulates this model.

\begin{subequations}
\begin{align}
    M_i &\sim \textrm{Bernoulli}\left(\theta_i\right)\label{eq:RQ3-m-1}\\
    \textrm{logit}\left(\theta_i\right) &= \alpha_{\mathrm{PRJ}\left[i\right]} + \beta_{\mathrm{PRJ}\left[i\right]} \times \textrm{Exec}_{i}+ \gamma_{\mathrm{PRJ}\left[i\right]} \times \textrm{Cover}_{i}\label{eq:RQ3-m-2}\\
    \alpha_{\mathrm{PRJ}\left[i\right]} &\sim \textrm{Normal}\left(0,1\right)\label{eq:RQ3-m-3}\\
    \beta_{\mathrm{PRJ}\left[i\right]} &\sim \textrm{Log{-}Normal}\left(0,1\right)\label{eq:RQ3-m-4}\\
    \gamma_{\mathrm{PRJ}\left[i\right]} &\sim \textrm{Log{-}Normal}\left(0,1\right)\label{eq:RQ3-m-5}\\
    \textrm{Exec}_{i} &\sim \textrm{Normal}\left(\mu_i, \sigma_i\right)\label{eq:RQ3-m-6}\\
    \mu_i &= \nu_{\mathrm{PRJ}\left[i\right]} + \lambda_{\mathrm{PRJ}\left[i\right]} \times \textrm{Cover}_{i}\label{eq:RQ3-m-7}\\
    \nu_{\mathrm{PRJ}\left[i\right]} &\sim \textrm{Normal}\left(0,1\right)\label{eq:RQ3-m-8}\\
    \lambda_{\mathrm{PRJ}\left[i\right]} &\sim \textrm{Normal}\left(0,1\right)\label{eq:RQ3-m-9}\\
    \sigma_i &\sim \textrm{Exponential}\left(1\right)\label{eq:RQ3-m-10}
\end{align}
\end{subequations}

The model comprises two parts. \cref{eq:RQ3-m-1} to \cref{eq:RQ3-m-5} are the first part, estimating the causal association between \textit{Exec} and the outcome concerned, which is the exact statistical model in RQ2. The second part is intended to estimate the path from \textit{Cover} to \textit{Exec} in \cref{fig:causal-dag}. 

We used Exponential distribution for the prior variance of \textit{Exec} since it is positive. It is a common choice for the variance as well~\citep{McElreath2020}\@. Other priors are similar to the ones defined in previous RQs.
Note that \cref{eq:RQ3-m-1} and \cref{eq:RQ1-m-1} are the same with the same reason we argued before for \cref{eq:RQ1-m-1} and \cref{eq:RQ2-m-1}.

This model enables us to answer some hypothetical questions. For example, we can estimate the influence of manipulating the variable \textit{Exec} (e.g., from {0} to {10000}) on a given mutant while holding the corresponding \textit{Cover} constant. We constructed the aforementioned model and compared it with the noncausal relationship obtained in RQ1 so that revealing the difference between the two.

\paragraph{RQ4: Causal association between \textit{Cover} and the outcome}\mbox{}\\
There is no back-door path from the \textit{Cover} variable to the mutant in \cref{fig:causal-dag}. Ergo we did not add \textit{Exec} to the statistical model so that it measures the causal association. It is tempting to include \textit{Exec} in the model, however, it is a common pitfall and leads to incorrect causal inference~\citep{McElreath2020}\@. The model specification comes next.

\begin{subequations}
\begin{align}
    M_i &\sim \textrm{Bernoulli}\left(\theta_i\right)\label{eq:RQ4-m-1}\\
    \textrm{logit}\left(\theta_i\right) &= \alpha_{\mathrm{PRJ}\left[i\right]} + \beta_{\mathrm{PRJ}\left[i\right]} \times \textrm{Cover}_{i}\label{eq:RQ4-m-2}\\
    \alpha_{\mathrm{PRJ}\left[i\right]} &\sim \textrm{Normal}\left(0,1\right)\label{eq:RQ4-m-3}\\
    \beta_{\mathrm{PRJ}\left[i\right]} &\sim \textrm{Log{-}Normal}\left(0,1\right)\label{eq:RQ4-m-4}
\end{align}
\end{subequations}

Conceptually, it is similar to the model defined in RQ1. However, the variable \textit{Exec} is replaced with \textit{Cover}. We use the same justification for the selection of priors in \cref{eq:RQ4-m-3} and \cref{eq:RQ4-m-4}, similar to the one in RQ1. 
Note that \cref{eq:RQ4-m-1} and \cref{eq:RQ1-m-1} are the same with the same reason we argued before for \cref{eq:RQ1-m-1} and \cref{eq:RQ2-m-1}.
The subtle difference between these two models demonstrates that inferring causal relationships can be challenging. The model RQ1 leads to noncausal association. Nonetheless, the model defined here is the correct one to estimate causal relationship.  

\subsection{Evaluation metrics}
We used posterior distribution of coefficients in the specified models in RQs as indicators of association (causal or noncausal) in the same line as \citet{McElreath2020}\@. We deliberately avoid using information criteria since this paper's focus is on the causal inference instead of predictive tasks. Information criteria are common when one tries to compare predictive performance of different models~\citep{McElreath2020}\@.

\subsection{Implementation details}
We used the Rethinking package~\citep{Rethinking2020}, an R library~\citep{R2020}, to implement the model specification~\citep{McElreath2020}. The Rethinking package is written on the top of Stan~\citep{Stan2021}\@, a probabilistic programming language. We exploited ggplot2, a graphical library, for plotting the figures in this article~\citep{Wickham2016}.

\section{Results}\label{sec:Results}

\subsection{Answers to RQ1}
\cref{tab:RQ1-answer} presents the average as well as credible intervals of {4000} samples drawn from the $\beta_{\mathrm{PRJ}\left[i\right]}$ posterior distributions for each project on the logit scale. Because these results are on the logit scale, they should be treated with caution while interpreting them~\citep{Fenton2012,Scholz2020}.

\textsc{wire} has the association with the mean of {2.79}. For each standard deviation increase in \textit{Exec} (or approximately {10}), expected odds of a mutant being killed (i.e., $\frac{\theta}{1 - \theta}$) increase by a factor of $\exp\left(2.79\right) = 16.28$. It is a significant influence. For example, for a mutant with random chance of being killed, increasing \textit{Exec} leads to roughly {5}\% probability of survival.
Having said that, it should not be considered as causal effect. As such, it does not mean that increase of \textit{Exec} by holding other factors constant has the same effect. 

The same interpretation is applicable to other projects. \textsc{Joda-time} and \textsc{assertj-core} have the association with the mean of {0.79} and {1.31} respectively. For each standard deviation increase in \textit{Exec} for those projects, expected odds of a mutant being killed increase by a factor of $\exp\left(0.79\right) = 2.20$ and $\exp\left(1.31\right) = 3.71$ respectively. Therefore, for a mutant with 50\% chance of being killed, increasing \textit{Exec} leads to roughly {70}\% and {80}\% probability of being killed respectively.

The credible interval provides probability interpretation for the results. As for \textsc{wire}, with the probability of {97.5}\% the odds of a mutant being killed increase at least by a factor of $\exp\left({2.63}\right) = {13.87}$ on average.

Different projects are of distinct associations. \textsc{lang} is a project with the least $\beta_{\mathrm{PRJ}\left[i\right]}$  magnitude among the subjects. The average {0.36} on the logit scale means for each standard deviation increase in \textit{Exec} (or roughly {15}), expected odds of a mutant being killed increase by a factor of $\exp\left({0.36}\right) = {1.43}$ suggesting that the association is weak. Increasing \textit{Exec} results to {60}\% or less chance of being killed for a mutant with a {50}\% of survival. Also, with the probability of {97.5}\% the odds of a mutant being killed increase at least by a factor of $\exp\left({0.32}\right) = {1.38}$ on average.

\begin{table}[ht]
    \centering
    \caption{Summary of {4000} samples from the $\beta_{\mathrm{PRJ}\left[i\right]}$ posterior distributions on the logit scale}\label{tab:RQ1-answer}
    \begin{tabular}{@{}lllll@{}}
      \toprule
      \textbf{Project}& \textbf{Mean}&   \textbf{SE}&   \textbf{{2.5}\%}&   \textbf{{97.5}\%}\\%
      \midrule
      \textsc{argparse4j}& {1.64}&   {0.05}&   {1.54}&   {1.75}\\
      \textsc{assertj-core}& {1.31}&   {0.03}&   {1.26}&  {1.37}\\
      \textsc{fess}& {2.18}&   {0.02}&   {2.14}&  {2.22}\\
      \textsc{joda-time}& {0.79}&   {0.02}&   {0.76}&   {0.83}\\
      \textsc{la4j}& {2.23}&   {0.04}&   {2.15}&   {2.31}\\
      \textsc{lang}& {0.36}&   {0.02}&   {0.32}&   {0.40}\\
      \textsc{msg}& {1.90}&   {0.04}&   {1.82}&   {1.98}\\
      \textsc{nodebox}& {3.00}&   {0.03}&   {2.94}&   {3.06}\\
      \textsc{opennlp}&  {1.26}&   {0.01}&   {1.25}& {1.28}\\
      \textsc{recast4j}& {1.49}&   {0.03}&   {1.44}&   {1.53}\\
      \textsc{uaa}& {1.64}&   {0.03}&   {1.58}&   {1.71}\\
      \textsc{wire}& {2.79}&   {0.08}&   {2.63}&   {2.95}\\
      \bottomrule
      \multicolumn{5}{@{}p{\linewidth}@{}}{SE refers to the standard error}\\
    \end{tabular}%
\end{table}

\begin{figure*}[ht]
  \centering
  \includegraphics[width=0.7\linewidth]{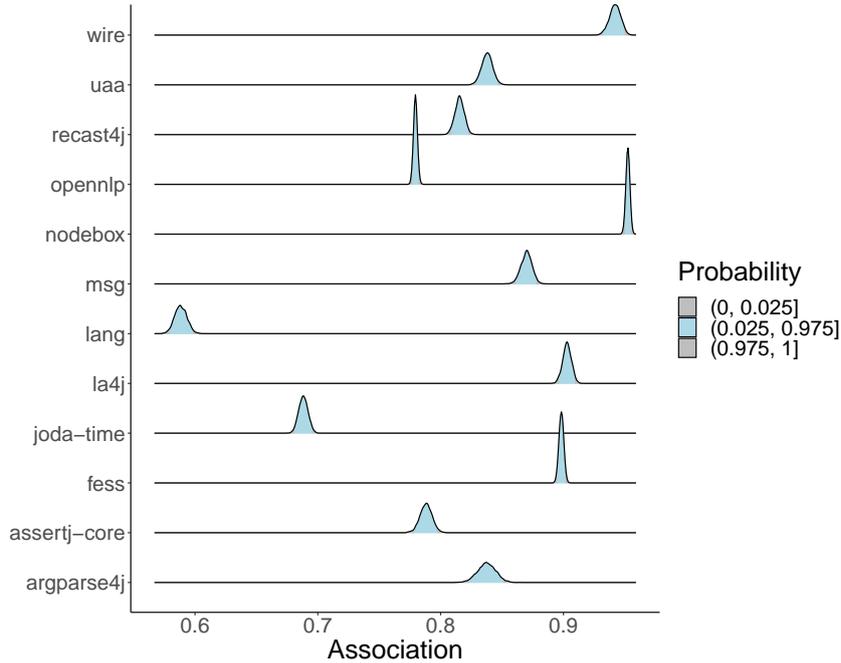}
  \caption{{4000} samples from the $\beta_{\mathrm{PRJ}\left[i\right]}$ posterior distributions on the output scale}\label{fig:RQ1-answer}
\end{figure*}

\cref{fig:RQ1-answer} illustrates the $\beta_{\mathrm{PRJ}\left[i\right]}$ posterior distributions on the probability scale, providing corroborating evidence for understanding the influence of \textit{Exec} on mutants.
In \cref{eq:RQ1-m-2}, $\beta_{\mathrm{PRJ}\left[i\right]}$ is the coefficient, which shows the importance of \textit{Exec}.
Unsurprisingly, larger projects have less uncertainty about the posterior distributions. For instance, \textsc{opennlp} possesses a narrow distribution, confirmed also by the standard error in \cref{tab:RQ1-answer}.

\textsc{nodebox} has the largest influence of \textit{Exec} on its mutants. For each standard deviation increase in \textit{Exec}, the expected probability of a mutant with random chance of being killed increases to ${95}$\%. The average {3.00} on the logit scale means for each standard deviation increase in \textit{Exec}, expected odds of a mutant being killed increases by a factor of $\exp\left({3.00}\right) = {20.09}$.

\subsection{Answers to RQ2}
\cref{tab:RQ2-answer} presents the causal effect of \textit{Exec} on the mutants. As opposed to \cref{tab:RQ1-answer}, some interesting dissimilarities are seen. \textsc{wire} is an example where for each standard deviation increase in \textit{Exec}, expected odds of a mutant being classified as killed increase by a factor of $\exp\left({1.84}\right) = 6.30$ (compared to {16.28} in RQ1).
As for \textsc{nodebox}, the average {1.27} on the logit scale means for each standard deviation increase in \textit{Exec}, expected odds of a mutant being killed increase by a factor  of $\exp\left({1.27}\right) = 3.56$ (compared to {20.09} in RQ1). This means the expected probability of a mutant with random chance of being killed increases to 80\% (compared to 95\% in RQ1).

Another appealing findings is in regard to \textsc{assertj-core}. For each standard deviation increase in \textit{Exec}, with the probability of ${97.5}\%$ the odds of a mutant being killed increase at most by a factor of $\exp\left({0.07}\right) = {1.07}$, which is small.

\begin{figure*}[ht]
  \centering
  \includegraphics[width=0.7\linewidth]{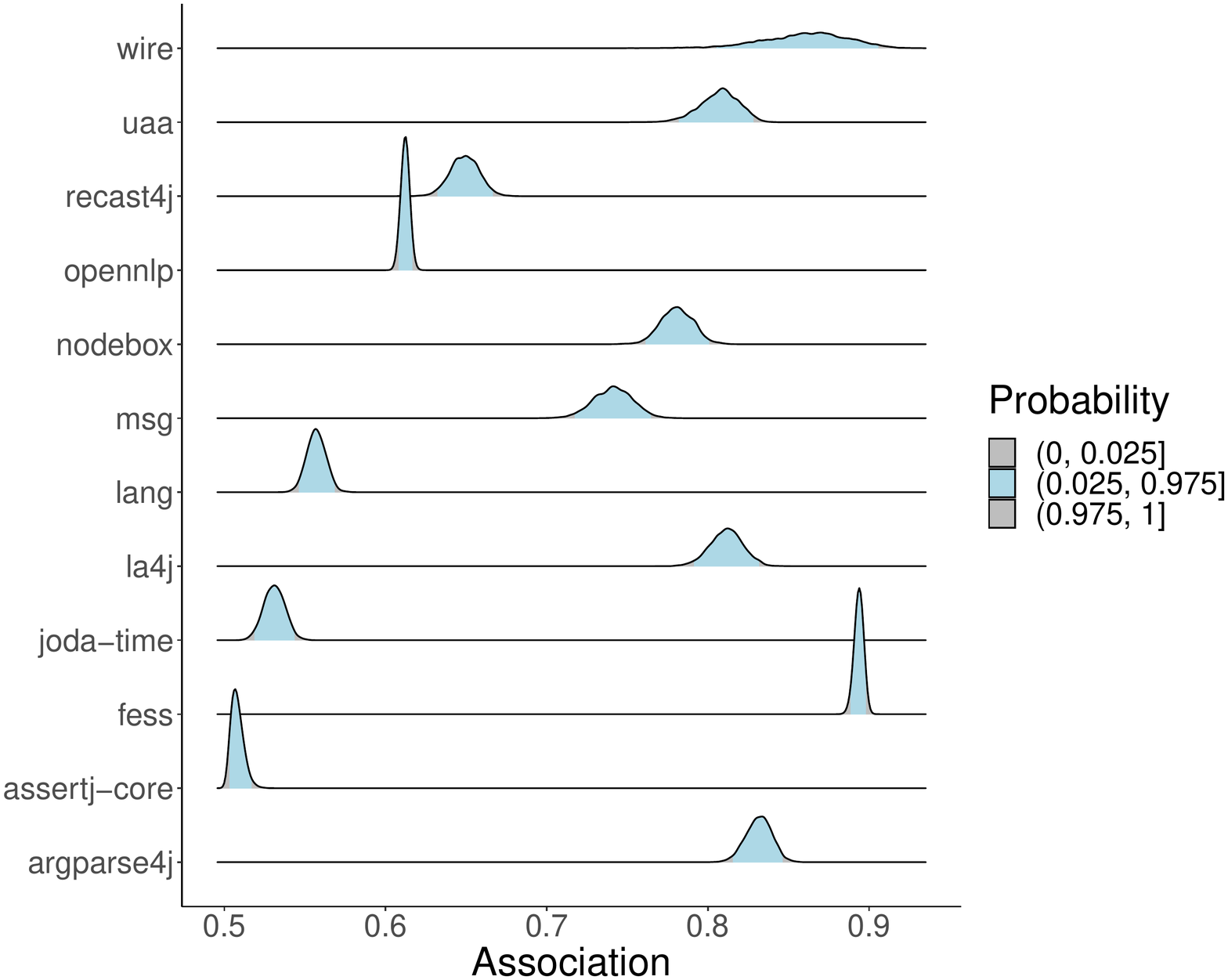}
  \caption{$\beta_{\mathrm{PRJ}\left[i\right]}$ posterior distributions on the output scale, showing the causal effect of \textit{Exec}}\label{fig:RQ2-answer}
\end{figure*}

The main reason for these decreases is the impact of \textit{Cover} which has a causal path to \textit{Exec} in \cref{fig:causal-dag}. However, these decreases vary between projects. \textit{Cover} exerts a mild influence on \textsc{fess}, i.e. {2.13} compared to {2.18} in RQ1. As the posterior probability distributions are at our disposal, we can quantify the uncertainties, drawing statistical inferences. \cref{fig:RQ2-answer} shows 
the $\beta_{\mathrm{PRJ}\left[i\right]}$ posterior distributions on the probability scale, which demonstrates the importance of \textit{Exec}. We can infer from this figure that \textsc{nodebox}, for instance, has a narrow posterior distribution, which is also confirmed by \cref{tab:RQ2-answer}.

\begin{table}[ht]
    \centering
    \caption{Summary of {4000} samples from the $\beta_{\mathrm{PRJ}\left[i\right]}$ posterior distributions on the logit scale, showing the causal effect of \textit{Exec}}\label{tab:RQ2-answer}
    \begin{tabular}{@{}lllll@{}}
      \toprule
      \textbf{Project}& \textbf{Mean}&   \textbf{SE}&   \textbf{{2.5}\%}&   \textbf{{97.5}\%}\\%
      \midrule
      \textsc{argparse4j}& {1.60}&   {0.06}&   {1.49}&   {1.70}\\
      \textsc{assertj-core}& {0.03}&   {0.01}&   {0.01}&  {0.07}\\
      \textsc{fess}& {2.13}&   {0.03}&   {2.07}&  {2.18}\\
      \textsc{joda-time}& {0.12}&   {0.03}&   {0.07}&   {0.17}\\
      \textsc{la4j}& {1.46}&   {0.07}&   {1.33}&   {1.59}\\
      \textsc{lang}& {0.23}&   {0.02}&   {0.18}&   {0.27}\\
      \textsc{msg}& {1.05}&   {0.06}&   {0.93}&   {1.17}\\
      \textsc{nodebox}& {1.27}&   {0.06}&   {1.16}&   {1.39}\\
      \textsc{opennlp}&  {0.46}&   {0.01}&   {0.44}& {0.48}\\
      \textsc{recast4j}& {0.62}&   {0.04}&   {0.54}&   {0.69}\\
      \textsc{uaa}& {1.43}&   {0.08}&   {1.27}&   {1.58}\\
      \textsc{wire}& {1.84}&   {0.22}&   {1.42}&   {2.26}\\
      \bottomrule
      \multicolumn{5}{@{}p{\linewidth}@{}}{SE refers to the standard error}\\
    \end{tabular}%
\end{table}

\cref{tab:RQ2-answer-comparison} reveals the inflation induced by the noncausal association obtained in RQ1. The results suggest the strong effect of \textit{Cover} on \textit{Exec} and outline the difference between causal and noncausal association. 

Some projects have been greatly affected, e.g. \textsc{nodebox} and \textsc{assertj-core}. As an illustration, the results for \textsc{assertj-core} is inflated with the probability of {97.5}\% at least by a factor of $\exp\left({1.25}\right)=3.5$, which is a massive difference. Nevertheless, it seems that some were less affected, e.g. \textsc{argparse4j} and \textsc{fess}. Ignoring \textit{Cover} leads to inflation of \textsc{argparse4j} by a factor of $\exp\left({0.04}\right)=1.04$ on average, which is negligible.

\begin{table}[ht]
    \centering
    \caption{Differences between posterior distributions of  $\beta_{\mathrm{PRJ}\left[i\right]}$ on the logit scale, obtained in RQ1 and RQ2}\label{tab:RQ2-answer-comparison}
    \begin{tabular}{@{}lllll@{}}
      \toprule
      \textbf{Project}& \textbf{Mean}&   \textbf{SE}&   \textbf{{2.5}\%}&   \textbf{{97.5}\%}\\%
      \midrule
      \textsc{argparse4j}& {0.04}&   {0.00}&   {0.04}&   {0.05}\\
      \textsc{assertj-core}& {1.28}&   {0.02}&   {1.25}&  {1.30}\\
      \textsc{fess}& {0.05}&   {0.01}&   {0.04}&  {0.07}\\
      \textsc{joda-time}& {0.67}&   {0.01}&   {0.65}&   {0.68}\\
      \textsc{la4j}& {0.77}&   {0.03}&   {0.71}&   {0.82}\\
      \textsc{lang}& {0.13}&   {0.00}&   {0.12}&   {0.13}\\
      \textsc{msg}& {0.85}&   {0.02}&   {0.80}&   {0.89}\\
      \textsc{nodebox}& {1.73}&   {0.03}&   {1.67}&   {1.78}\\
      \textsc{opennlp}&  {0.81}&   {0.00}&   {0.80}& {0.81}\\
      \textsc{recast4j}& {0.87}&   {0.01}&   {0.84}&   {0.90}\\
      \textsc{uaa}& {0.21}&   {0.04}&   {0.13}&   {0.31}\\
      \textsc{wire}& {0.95}&   {0.13}&   {0.69}&   {1.21}\\
      \bottomrule
      \multicolumn{5}{@{}p{\linewidth}@{}}{SE refers to the standard error}\\
    \end{tabular}%
\end{table}

\subsection{Answers to RQ3}

\begin{figure}
    \centering
  \subfloat[\textsc{nodebox}]{%
    \includegraphics[width=.45\linewidth]{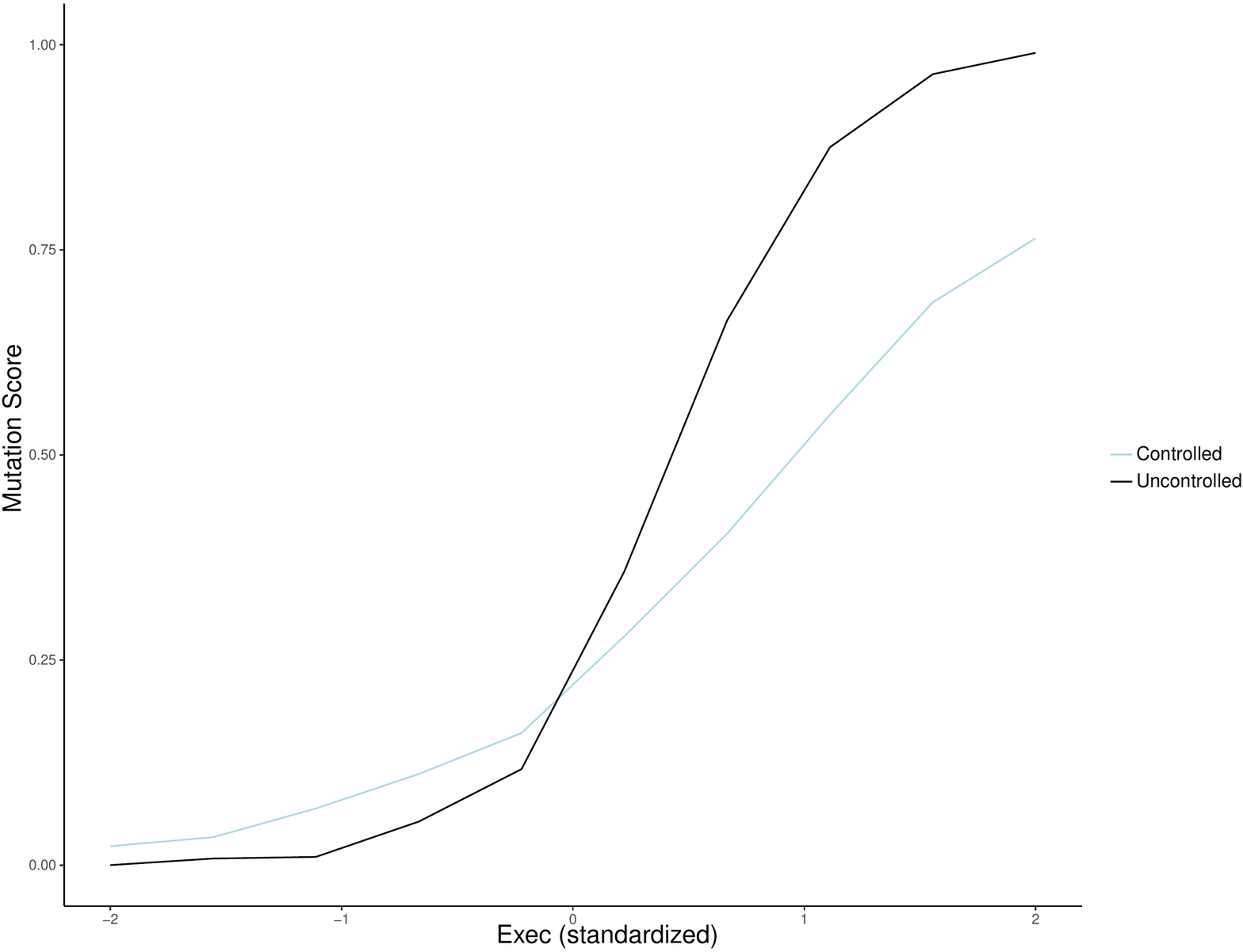}}\hfill
      \subfloat[\textsc{assertj-core}]{%
    \includegraphics[width=0.45\linewidth]{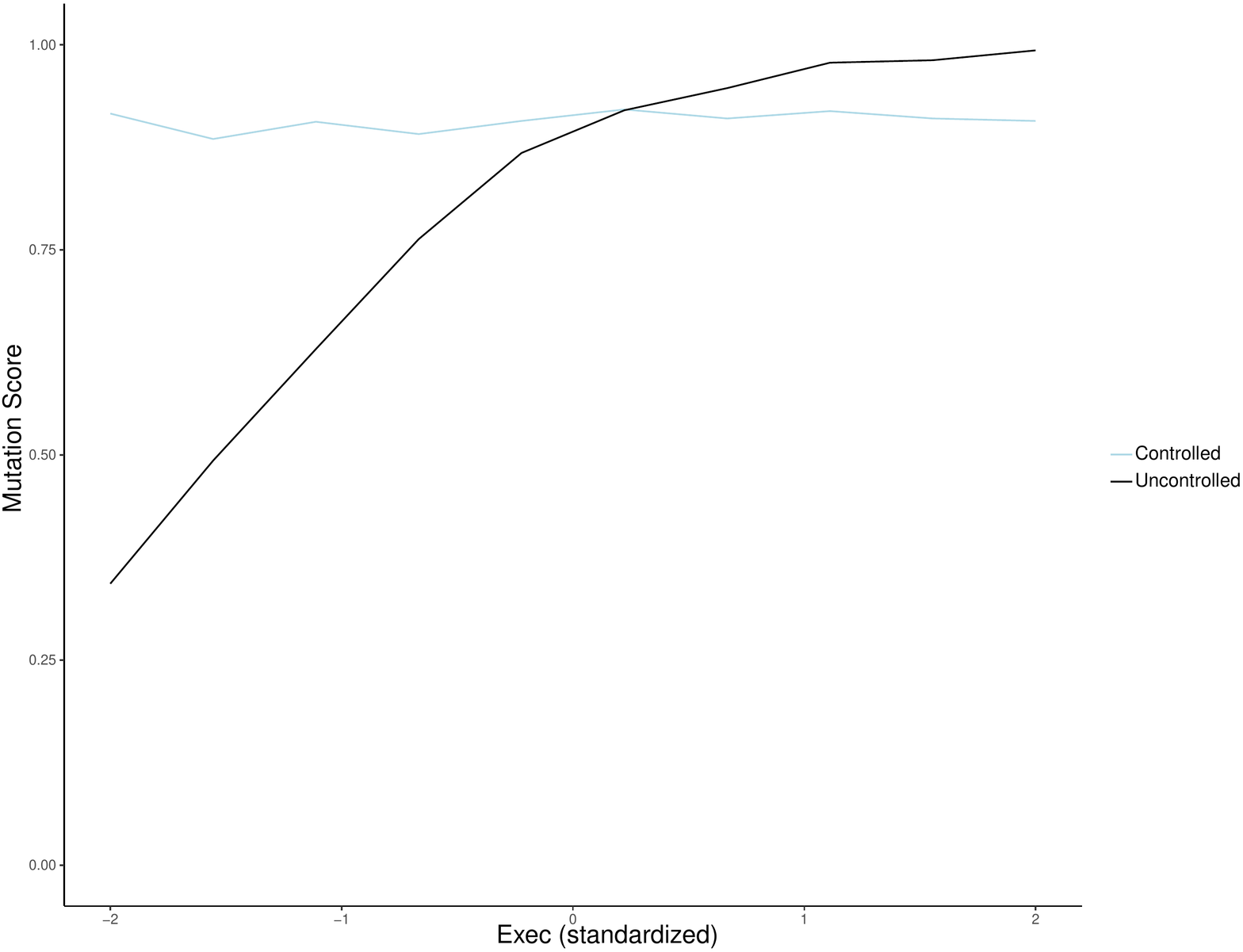}}\hfill
    \caption{Counterfactual plots in which \textit{Cover} is set to average for \textsc{nodebox} and \textsc{assertj-core}}
    \label{fig:counterfact}
\end{figure}

\cref{fig:counterfact} demonstrates the counterfactual plots in which the \textit{Cover} variable is set to average for the \textsc{nodebox} and \textsc{asssertj-core} projects. Interested readers are referred to the appendix for counterfactual plots of other projects. The light lines are the causal influence of changing \textit{Cover}, as opposed to the dark ones where describe the pitfall explained in RQ1. 

These plots provide the tools which enable us to answer hypothetical questions. As an example, holding \textit{Cover} constant, for two standard deviation increase in \textit{Exec}, the mutation score of \textsc{nodebox} increases to {0.75}. Having said that, the noncausal association implies that the mutation score would be approximately {1}, which is incorrect.

There is hardly any change in the mutation score of \textsc{assertj-core} by increasing/decreasing \textit{Exec}, as against the noncausal association inducing the mutation score of less than {0.40} for two standard deviations decrease in \textit{Exec}. This finding is confirmed before in \cref{tab:RQ2-answer}.

The counterfactual plots reveal that \textit{Exec} is an important causal factor for mutant execution results, although small compared to the noncausal association.

\subsection{Answers to RQ4}
The causal association between \textit{Cover} and mutants is detailed in \cref{tab:RQ4-answer}. \textsc{nodebox} possesses the largest effect of \textit{Cover}. For each standard deviation increase in \textit{Cover} (i.e., about {3}), the expected odds of a mutant being killed are multiplied by $\exp\left({3.28}\right) = {26.58}$. As such, the probability of a mutant with {50}\% chance of being classified as killed increases to {96}\%. 

\begin{figure*}[ht]
  \centering
  \includegraphics[width=0.7\linewidth]{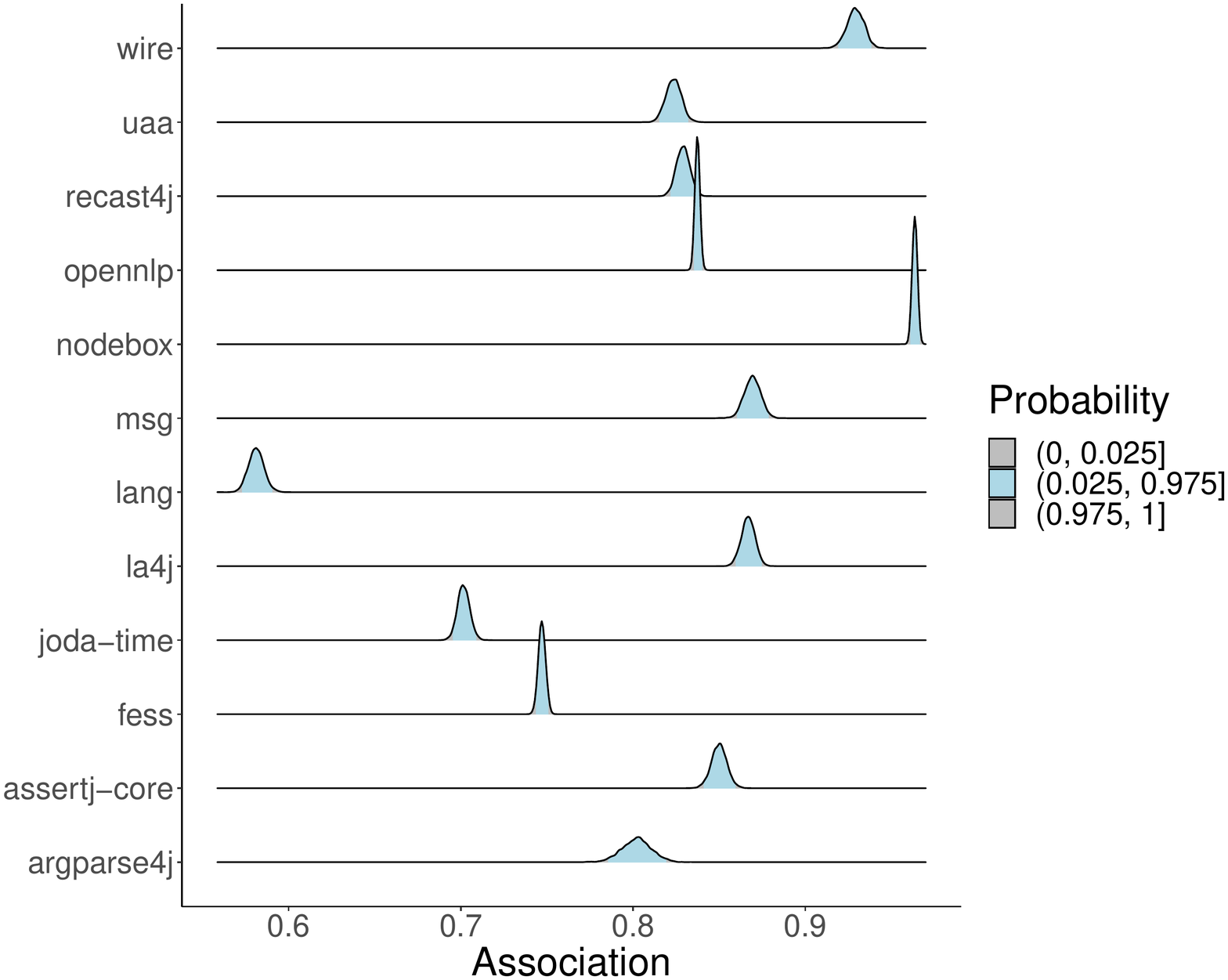}
  \caption{$\beta_{\mathrm{PRJ}\left[i\right]}$ posterior distributions on the output scale, showing the causal association of \textit{Cover}}\label{fig:RQ4-answer}
\end{figure*}

\textsc{lang}, on the other hand, has been barely affected. For each standard deviation increase in \textit{Cover}, the odds of a mutant being classified as killed increase by a factor of $\exp\left({0.33}\right) = 1.39$ on average. Ergo, the probability of a mutant with {50}\% chance of being killed increases to {58}\%. This fact is outlined in \cref{fig:RQ4-answer}. Except for \textsc{lang}, odds of a mutant being killed for other projects have been substantially affected at least by a factor of $3$.

\begin{table}[ht]
    \centering
    \caption{Summary of {4000} samples from the $\beta_{\mathrm{PRJ}\left[i\right]}$ posterior distributions on the logit scale, showing the causal association of \textit{Cover}}\label{tab:RQ4-answer}
    \begin{tabular}{@{}lllll@{}}
      \toprule
      \textbf{Project}& \textbf{Mean}&   \textbf{SE}&   \textbf{{2.5}\%}&   \textbf{{97.5}\%}\\%
      \midrule
      \textsc{argparse4j}& {1.40}&   {0.05}&   {1.30}&   {1.51}\\
      \textsc{assertj-core}& {1.73}&   {0.04}&   {1.67}&  {1.81}\\
      \textsc{fess}& {1.08}&   {0.01}&   {1.06}&  {1.10}\\
      \textsc{joda-time}& {0.85}&   {0.02}&   {0.82}&   {0.89}\\
      \textsc{la4j}& {1.87}&   {0.03}&   {1.81}&   {1.94}\\
      \textsc{lang}& {0.33}&   {0.02}&   {0.29}&   {0.36}\\
      \textsc{msg}& {1.90}&   {0.04}&   {1.82}&   {1.98}\\
      \textsc{nodebox}& {3.28}&   {0.04}&   {3.20}&   {3.35}\\
      \textsc{opennlp}&  {1.64}&   {0.01}&   {1.62}& {1.66}\\
      \textsc{recast4j}& {1.58}&   {0.03}&   {1.53}&   {1.63}\\
      \textsc{uaa}& {1.54}&   {0.03}&   {1.48}&   {1.60}\\
      \textsc{wire}& {2.58}&   {0.08}&   {2.43}&   {2.73}\\
      \bottomrule
      \multicolumn{5}{@{}p{\linewidth}@{}}{SE refers to the standard error}\\
    \end{tabular}%
\end{table}

The question arises as to whether \textit{Cover} in comparison to \textit{Exec} has more causal effect on mutants or not. \cref{tab:RQ4-answer-comparison} contrasts \textit{Cover} with \textit{Exec} in terms of causal association, on the logit scale. Except for \textsc{argparse4j} and \textsc{fess}, the influence of \textit{Cover} is larger than \textit{Exec}, suggesting than both variables are important causally in killing mutants. However, it seems that overall \textit{Cover} is of more causal association as against \textit{Exec}. Note that the \textsc{fess} mutation score is just {0.05} highlighting the fact that many mutants are not executed and not covered.

\begin{table}[htpb]
    \centering
    \caption{The difference between causal association of \textit{Cover} and \textit{Exec}.}\label{tab:RQ4-answer-comparison}
    \begin{tabular}{@{}lllll@{}}
      \toprule
      \textbf{Project}& \textbf{Mean}&   \textbf{SE}&   \textbf{{2.5}\%}&   \textbf{{97.5}\%}\\%
      \midrule
      \textsc{argparse4j}& {-0.20}&   {0.00}&   {-0.20}&   {-0.19}\\
      \textsc{assertj-core}& {1.70}&   {0.02}&   {1.65}&  {1.74}\\
      \textsc{fess}& {-1.05}&   {0.02}&   {-1.07}&  {-1.01}\\
      \textsc{joda-time}& {0.73}&   {0.01}&   {0.72}&   {0.75}\\
      \textsc{la4j}& {0.41}&   {0.03}&   {0.35}&   {0.48}\\
      \textsc{lang}& {0.10}&   {0.00}&   {0.09}&   {0.11}\\
      \textsc{msg}& {0.84}&   {0.02}&   {0.80}&   {0.89}\\
      \textsc{nodebox}& {2.00}&   {0.02}&   {1.96}&   {2.05}\\
      \textsc{opennlp}&  {1.18}&   {0.00}&   {1.18}& {1.18}\\
      \textsc{recast4j}& {0.96}&   {0.01}&   {0.94}&   {0.99}\\
      \textsc{uaa}& {0.11}&   {0.05}&   {0.03}&   {0.21}\\
      \textsc{wire}& {0.74}&   {0.14}&   {0.47}&   {1.01}\\
      \bottomrule
      \multicolumn{5}{@{}p{\linewidth}@{}}{SE refers to the standard error}\\
    \end{tabular}%
\end{table}

\section{Discussion}\label{sec:Discussion}
In this section, we examine the limitation of this study and explain its impact on both researchers and software engineers.

\begin{figure}[ht]
    \centering
    \includegraphics[width=0.7\linewidth]{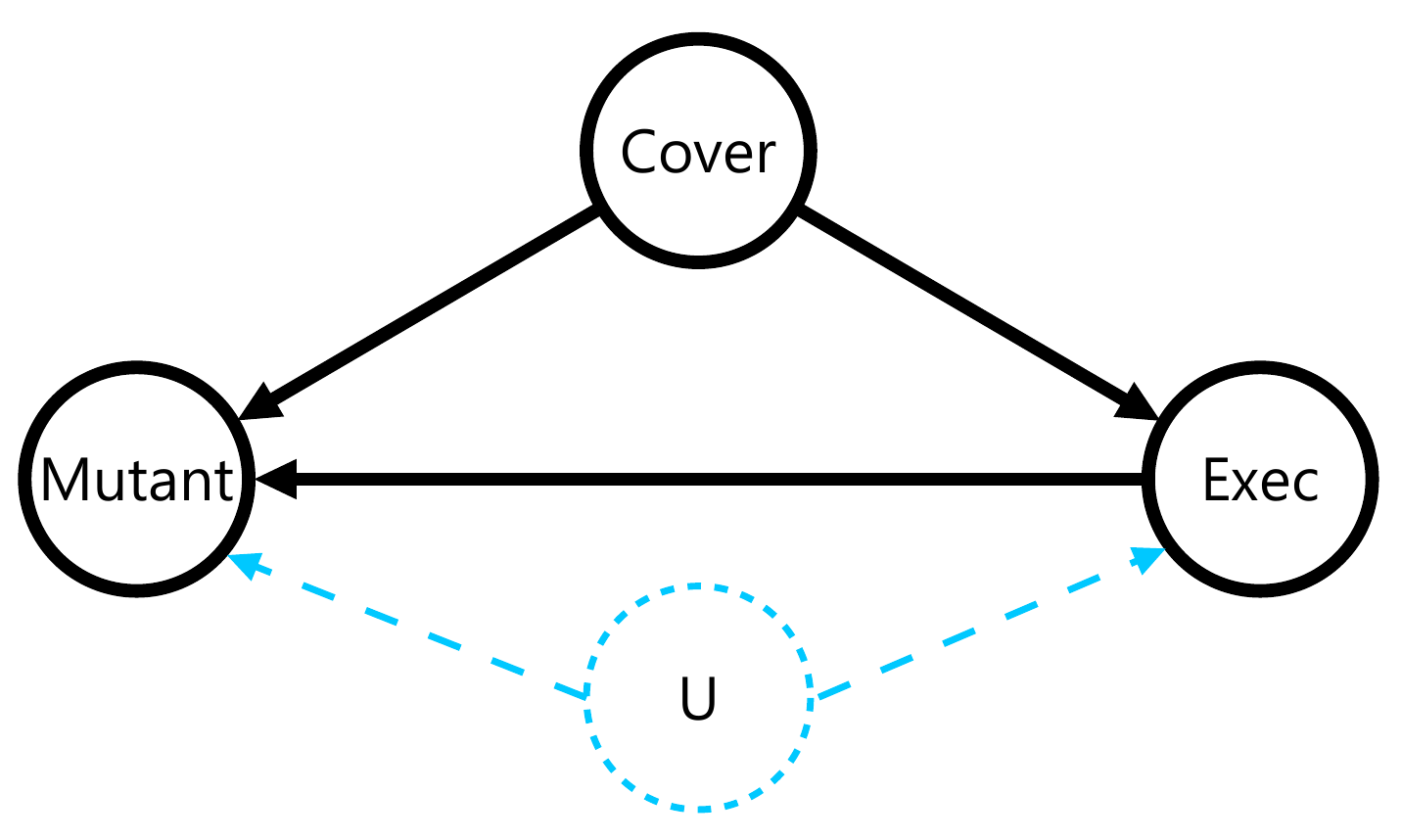}
    \caption{Unobserved confounding covariate}
    \label{fig:udag}
\end{figure}

Since this study is observational, unobserved covariates are likely to exist. If an unobserved covariate, \textit{U}, arises as a confounding variable (e.g, \cref{fig:udag}), the causal inference changes because of the new back-door path from \textit{Exec} to the outcome, i.e., $\textrm{Exec} \leftarrow U \rightarrow \textrm{Mutant}$. Although this limitation exists, the main finding  still holds (i.e., \cref{tab:RQ2-answer-comparison}) due to the fact that the confounding covariates increase the difference between the associations obtained in RQ1 and the ones in RQ2.
In general, it is infeasible to know, in observational studies, whether unobserved confounding variables exist~\citep{Guo2020,Neal2021}\@. Note that metrics such as the number of assertions, the code complexity, and the code quality metrics are not confounders as they do not create back-door paths from \textit{Exec}/\textit{Cover} to the outcome.

Bayesian statistics is a superb tool, supporting causal inference in defining multi-level GLMs. The combination of Bayesian statistics and causal inference is commonplace in literature~\citep{Gelman2020,McElreath2020}\@. In this study, we analyzed a simple issue regarding test efficacy. Researchers could follow the same steps to draw causal inferences from intended observed data~\citep{McElreath2020}\@:
\begin{enumerate}
\item \textbf{Define the outcome of interest and covariates.} It includes the type of variables (i.e., nominal, ordinal, interval, or ratio), the unit level of measures, and the range of possible values~\citep{Shull2008}.
\item \textbf{Draw a causal graph}. It represents the causal relationship between variables. In particular, each edge either indicates the causal association between two covariates or the causal association from a covariate to the outcome~\citep{McElreath2020}\@.
\item \textbf{Model the problem using GLMs, conditioning on covariates obtained in the back-door criterion.} The model definition comprises the likelihood, the \textit{link function} (i.e., the mapping function from covariates to the unknown parameters), priors, and justification for the selected priors (e.g., prior predictive simulations or posterior checks)~\citep{McElreath2020}.\@
\item \textbf{Measure the association between the covariate of interest and the outcome.} Using coefficients as the representation of relationship strength is commonplace~\citep{McElreath2020}\@. Apart from the expected value of coefficients, the posterior distribution and its positions (i.e., above or below zero) are of value.
\end{enumerate}

Software developers, quality assurance teams, team leaders, and project managers can benefit from causal inference. Developers could exploit explainable test effectiveness to design and write better test cases, which lead to discovering more bugs hidden in software. Managers could use insights from these results to plan, acquire, monitor, and control the time and budget of their development teams. For example, if in a project $Cover$ is of more causal association as against $Exec$, it suggests that a manager should strictly monitor and control $Cover$ values.

\section{Threats to validity}\label{sec:Threats}
We discuss threats to the validity of this paper, namely construct validity, internal validity and external validity~\citep{Feldt2010}\@. The reliability is also elaborated here~\citep{Herbold2021}\@.
\subsection{Construct validity}
We used mutants as a substitute for faults and exploited mutation score as operationalization of test efficacy. Whether mutants are apt for assessing test effectiveness is, in part, a matter of debate in literature. Majority have suggested that mutants are a good proxy for faults~\citep{Andrews2005,Andrews2006,Zhang2015}\@, and using mutants to assess test efficacy has been seen in many studies~\citep{Aghamohammadi2021,Gligoric2015,Gopinath2014,Zhang2015}\@. \citet{Papadakis2018}, however, argued that when the test size, as a confounder, is controlled, the correlation between test efficacy and mutation score becomes weak. 
\subsection{Internal validity}
As discussed in \cref{sec:Discussion}, the potential risk of an unobserved confounding covariate exists, threatening the internal validity of the findings~\citep{Kashner2019}\@. The choice of priors is another threat. However, we used weakly informative priors, prior predictive simulations, and posterior checks to mitigate this threat. To measure statistical dependence, multi-level GLM was applied. It is not mandatory to model the scientific problem by GLM~\citep{McElreath2020}\@. Nevertheless, multi-level GLM is a well-known and common approach in causal inference.
We did not manually eliminate equivalent mutants (except those being automatically eliminated by the mutation testing tool) which in turn affect the posterior distributions due to the fact that they are classified as alive. Excluding equivalent mutants, in general, is an undecidable problem~\citep{Frankl1998,Koster2007,Offutt1996}.
Another threat might be coarse-grained tests in which multiple checks occur and could be split into multiple fine-grained test cases. This can affect the results since the value of \textit{Cover} would increase due to splitting tests.

\subsection{External validity}
We employed {12} Java projects in this study and the findings may not be generalized to other programming languages such as Python or JavaScript. 
\change{For example, Python and JavaScript are dynamically-typed programming languages, which are different from statically-typed programming languages such as Java. In those languages, type checking happens at the execution time. Therefore, the kinds of faults in Python and JavaScript may be different than in Java. In our future work, we plan to create a dataset of faults accompanied by metrics such as \textit{Exec} and \textit{Cover}. That dataset could be a foundation for employing causal inference in an observational study.}
Moreover, selected projects were open source and there is a chance that the results do not hold true for closed source programs. We plan to collaborate with private companies to access closed source program and seek any difference in findings.

\subsection{Reliability}
Reliability refers to the consistency of measurements~\citep{Mohamad2015}\@. In the results section, we reported standard errors for the samples drawn from posterior distributions (e.g., \cref{tab:RQ1-answer}). The means of posterior distributions are greater than standard errors at least by a factor of two, e.g., for \textsc{uaa} the ratio of the mean to the standard error is $\frac{{0.11}}{{0.05}} > 2$ (see \cref{tab:RQ4-answer-comparison}). It suggests that the results provided in \cref{sec:Results} are reliable.

\section{Conclusions}\label{sec:Conclusions}
In this research, we showed that the naive association between \textit{Exec} and mutant status differs from the causal association one, highlighting the fact that to infer causal inference in an observational study, researchers should focus on the causal DAG and condition appropriately to block noncausal paths. We conducted research, measuring the noncausal association between \textit{Exec} and the outcome, the causal relationship between \textit{Exec} and the mutant execution results, and the causal association between \textit{Cover} and the outcome of interest. 

The results show that noncausal and causal association between \textit{Exec} and the outcome are different statistically. To determine the causal effect of \textit{Exec}, one should control for the \textit{Cover} variable. Nevertheless, when it comes to measuring the causal effect of \textit{Cover} on mutants, one should not condition on \textit{Exec}. The reason goes back to the back-door criterion.

 The findings provided in this paper highlight that even in the simple statistical model with only two independent variables it is easy to make an inaccurate inference. And researchers should treat their interpretations with caution since causal and noncausal associations are different.

\section*{Acknowledgments}
We would like to thank Maliheh Izadi and Mahtab Nejati for their beneficial comments.

\bibliography{ref}

\appendix

\section{Counterfactual Plots}
\begin{figure*}[ht]
\centering
  \subfloat[\textsc{argparse4j}]{%
    \includegraphics[width=0.48\linewidth]{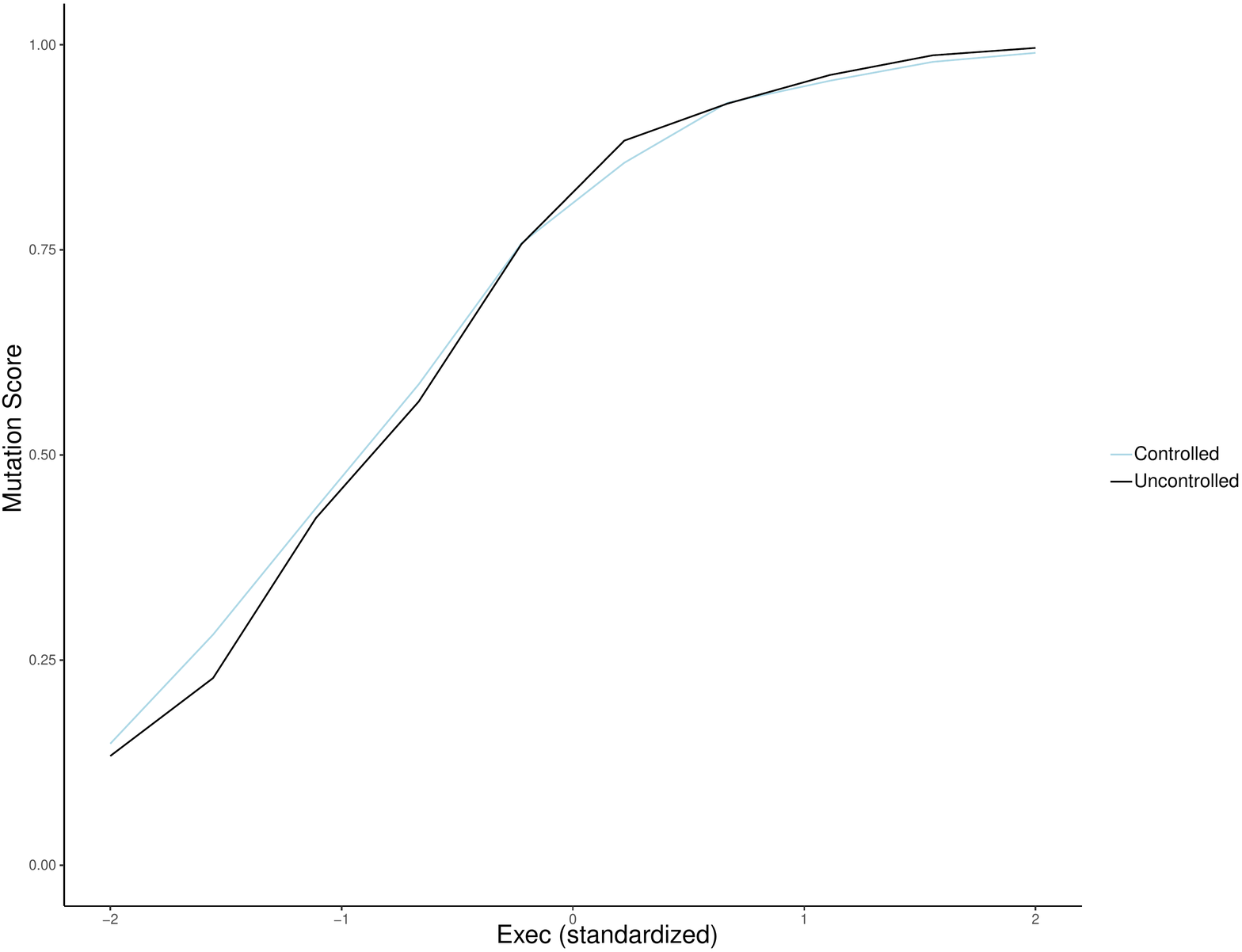}}\hfill
  \subfloat[\textsc{fess}]{%
    \includegraphics[width=0.48\linewidth]{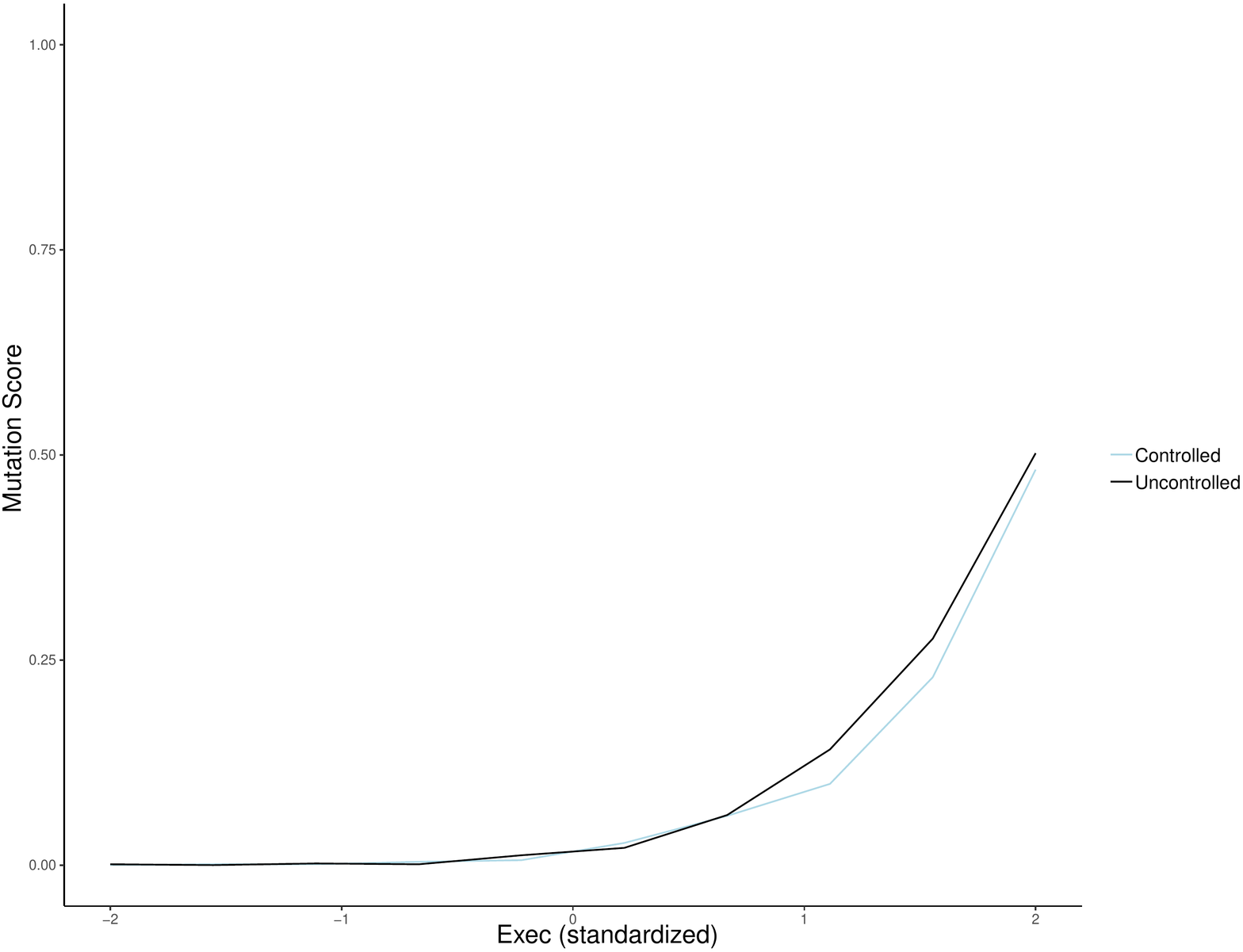}}\\
  \end{figure*}
  \begin{figure*}[ht]
  \centering
    \ContinuedFloat
         \subfloat[\textsc{joda-time}]{%
    \includegraphics[width=.48\linewidth]{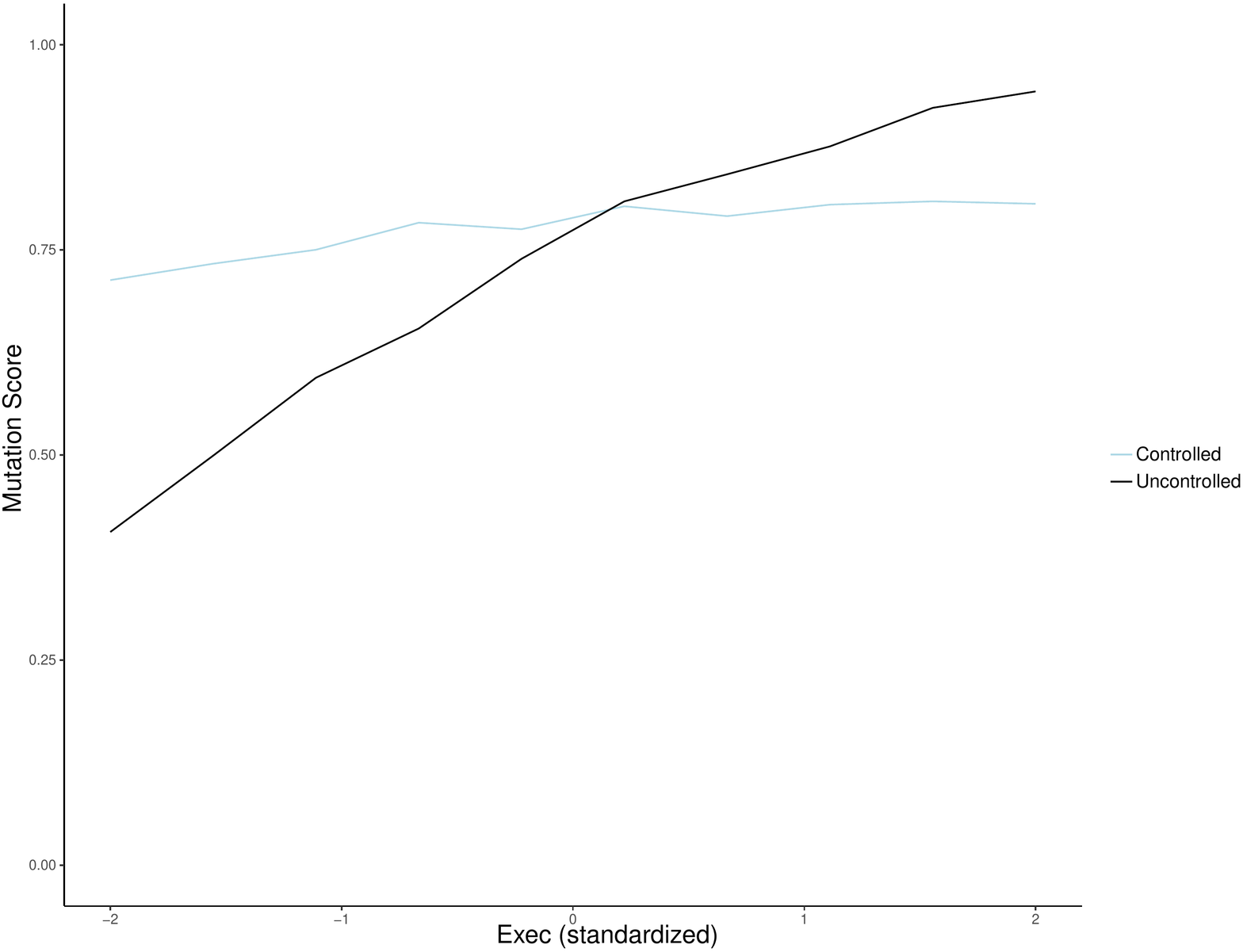}}\hfill
  \subfloat[\textsc{la4j}]{%
    \includegraphics[width=.48\linewidth]{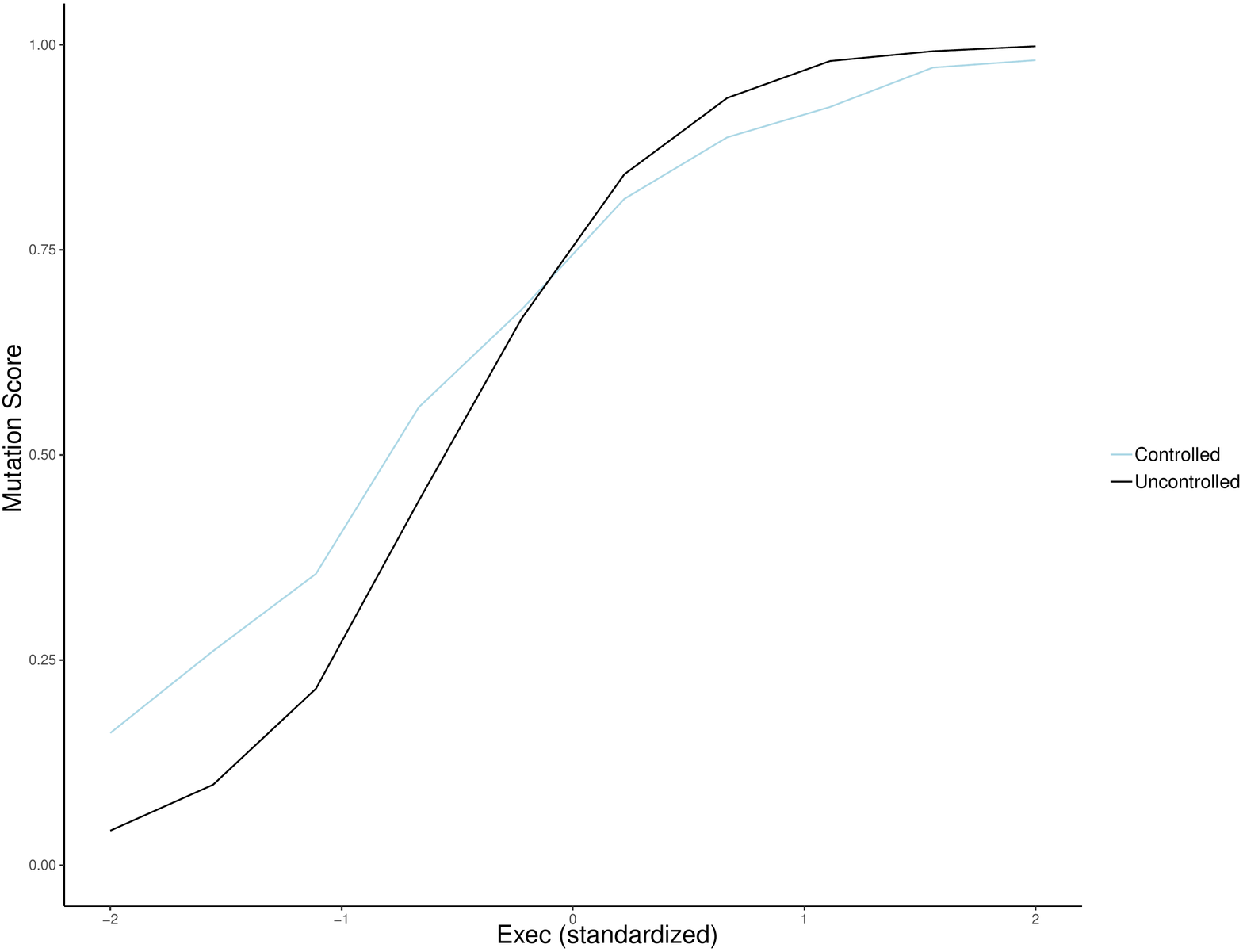}}\\
\end{figure*}
\begin{figure*}[ht]
  \centering
    \ContinuedFloat
  \subfloat[\textsc{lang}]{%
    \includegraphics[width=.48\linewidth]{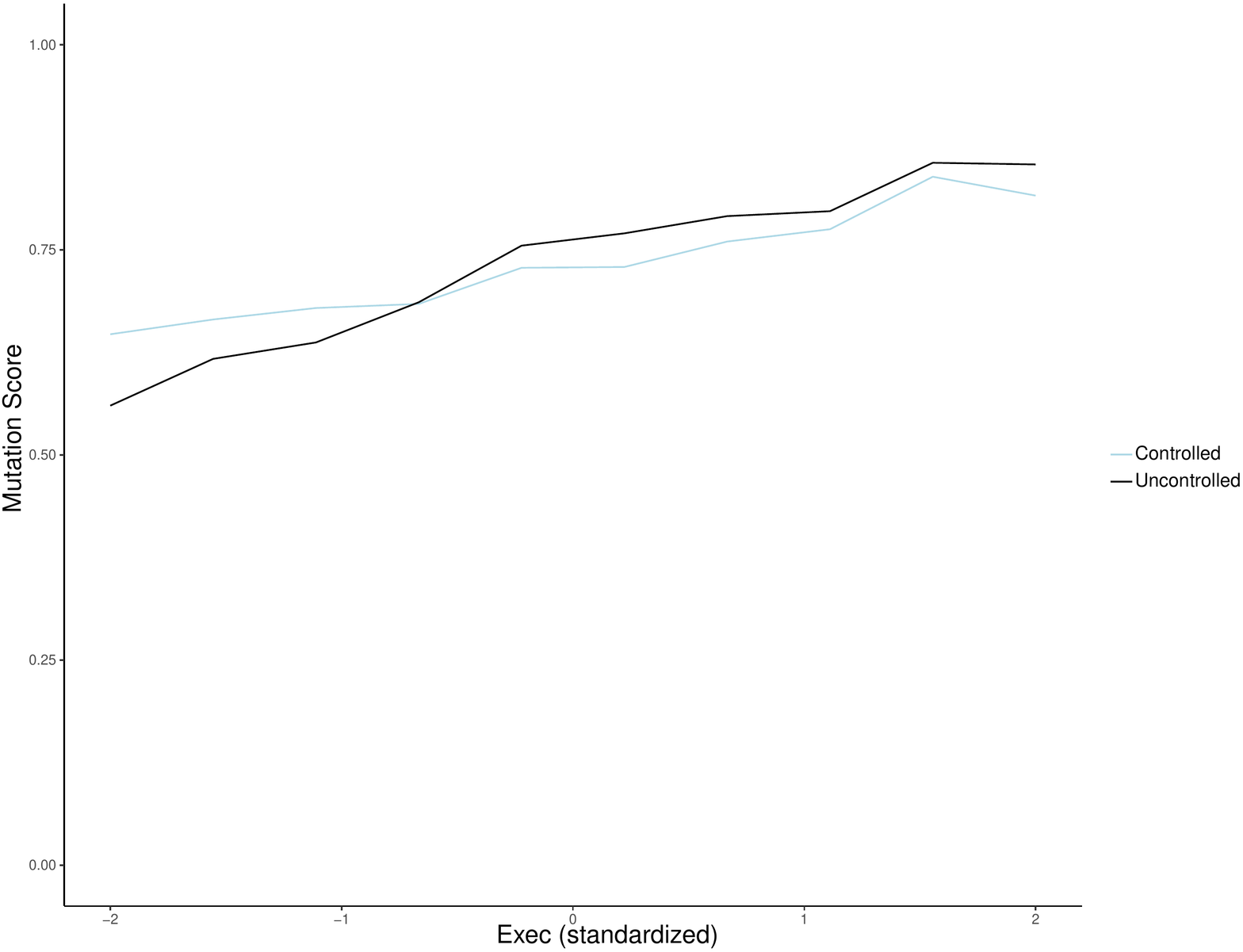}}\hfill
      \subfloat[\textsc{msg}]{%
    \includegraphics[width=.48\linewidth]{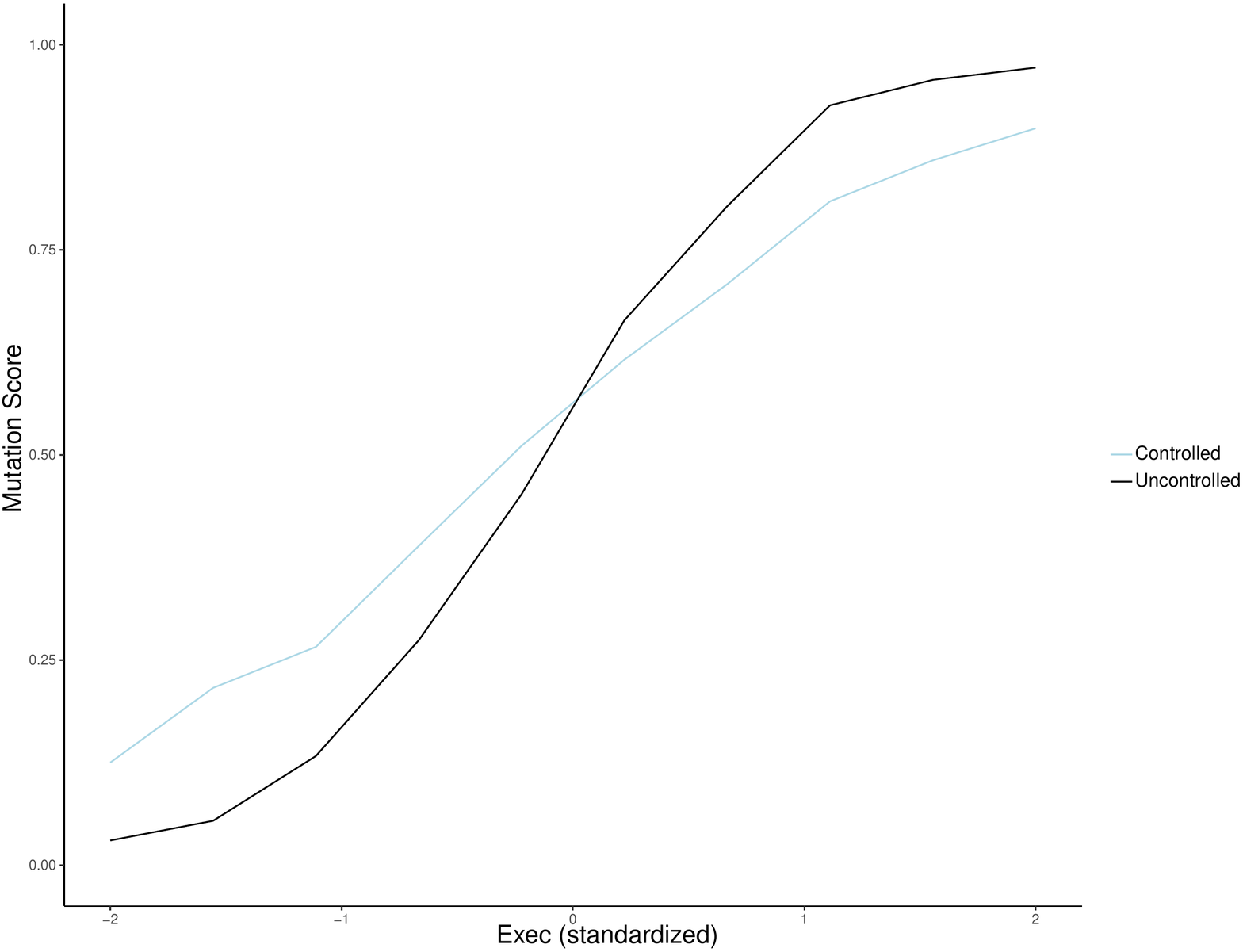}}\\
\end{figure*}

\begin{figure*}[ht]
\centering
    \ContinuedFloat
  \subfloat[\textsc{opennlp}]{%
    \includegraphics[width=.48\linewidth]{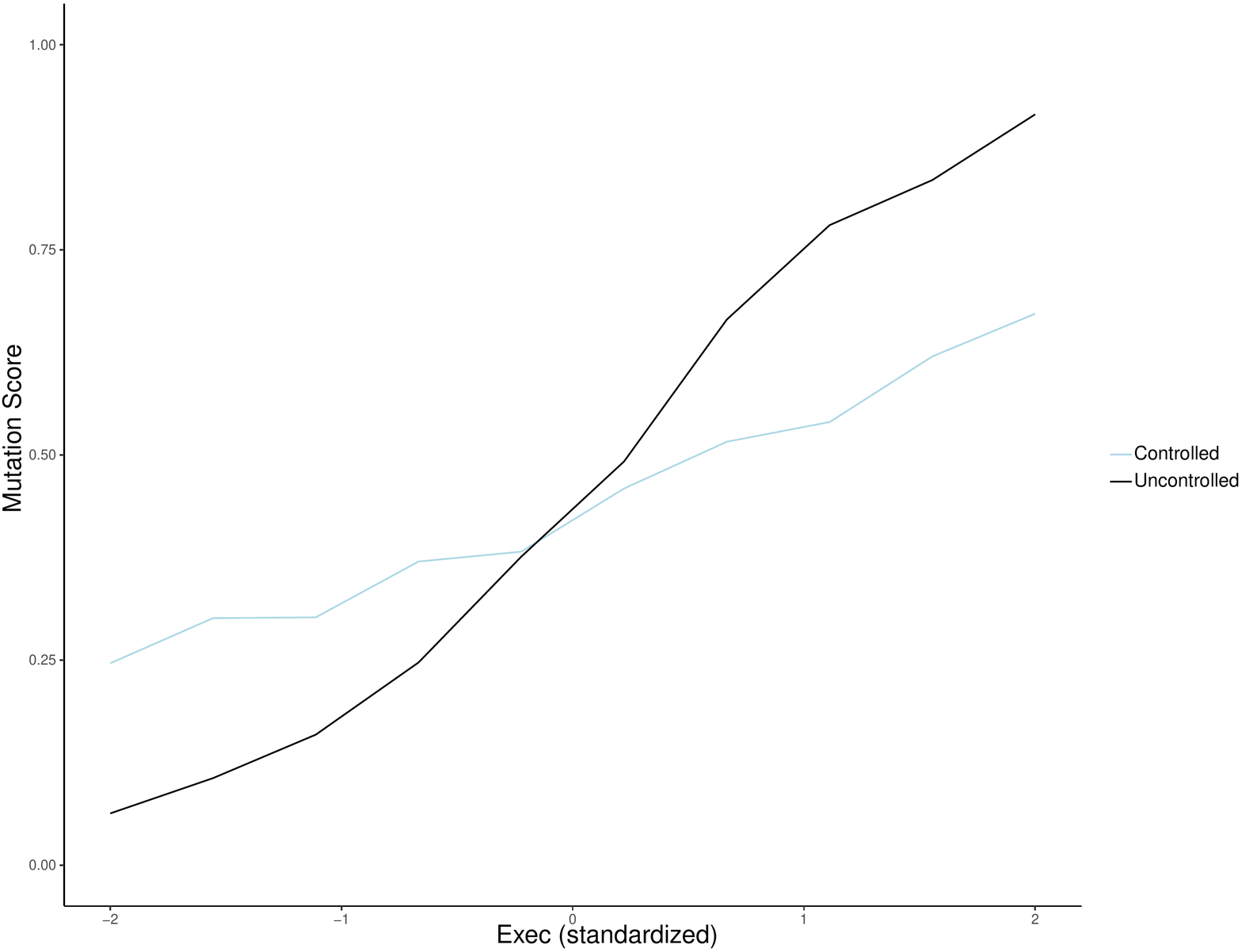}}\hfill
     \subfloat[\textsc{recast4j}]{%
    \includegraphics[width=.48\linewidth]{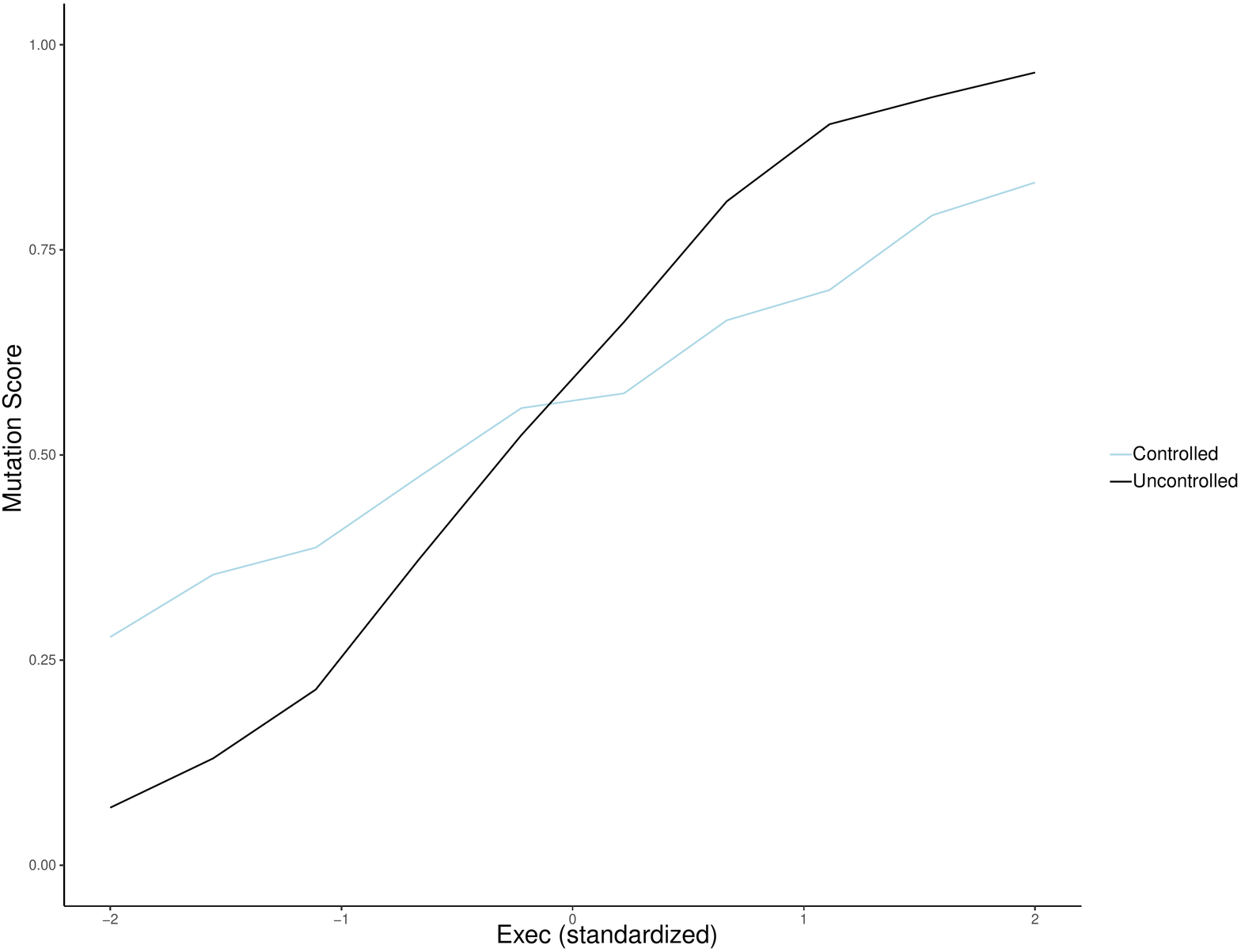}}\\
  \end{figure*}

 \begin{figure*}[ht]
\centering
    \ContinuedFloat
   \subfloat[\textsc{uaa}]{%
    \includegraphics[width=.48\linewidth]{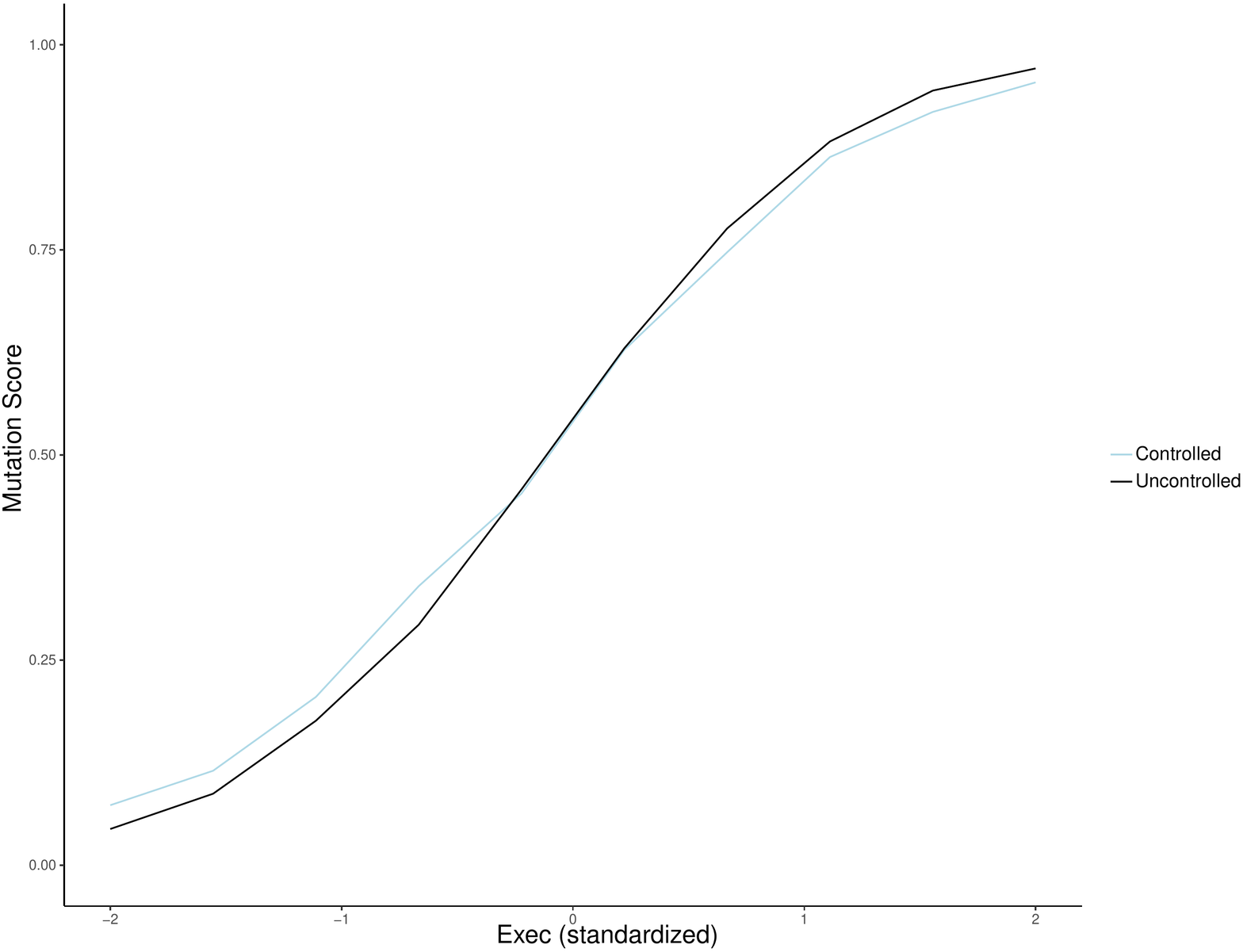}}\hfill
  \subfloat[\textsc{wire}]{%
    \includegraphics[width=.48\linewidth]{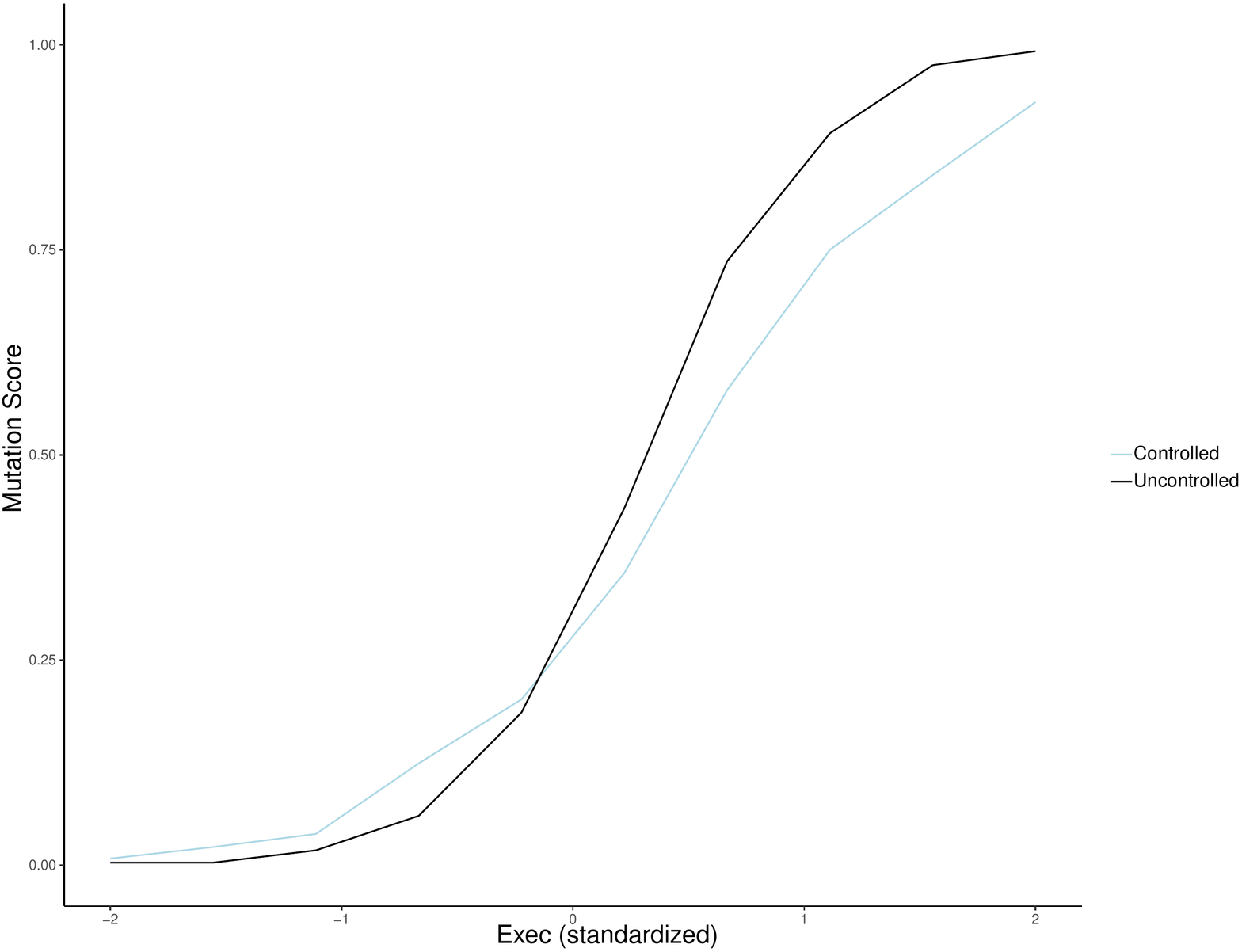}}\\
  \caption{Counterfactual plots in which \textit{Cover} is set to average}\label{fig:cfp}
\end{figure*}

\end{document}